\documentclass[%
 reprint,
 superscriptaddress,
 amsmath,
 amssymb,
 aps,
 prx,
 10pt
]{revtex4-2}

\usepackage{physics}
\usepackage[dvipsnames]{xcolor} % Colors
\usepackage[hidelinks]{hyperref} % Hyperlinks
\usepackage{hyperref}
\usepackage{placeins}
\usepackage{afterpage}
\usepackage{mathtools}
\usepackage{verbatim}
\usepackage{graphicx}% Include figure files
\usepackage{dcolumn}% Align table columns on decimal point
\usepackage{bm}% bold math
\usepackage{ulem}
\usepackage{tabularray}
\usepackage{stmaryrd} 
\usepackage{booktabs}
\usepackage{tabularx}
\usepackage{multirow}
\usepackage{adjustbox}
\usepackage{eso-pic}
\usepackage{graphicx}
\usepackage{siunitx}

\usepackage[group-minimum-digits=0,table-figures-decimal=0,table-number-alignment=center]{siunitx}
\usepackage[indent=10pt, skip=2pt plus1pt]{parskip} % Change paragraph spacing
\usepackage{makeidx}
\usepackage{lipsum} % Lorem ipsum

\hypersetup{
    colorlinks=true,
    linkcolor=blue,
    filecolor=magenta,      
    urlcolor=cyan,
    pdftitle={main},
    pdfpagemode=FullScreen,
}

\newcommand{\tetris}{\includegraphics[scale=0.043]{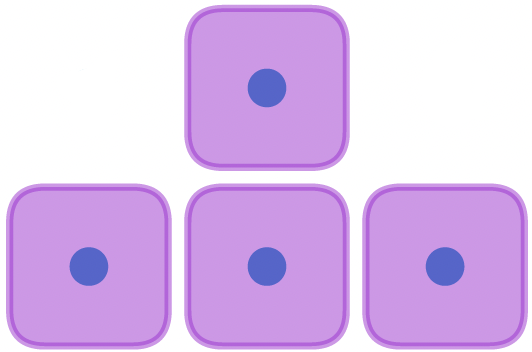}}

\newcommand{\etoile}{\includegraphics[scale=0.3]{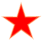}}

\makeindex

\begin{document}

\begin{comment} %to add the "Draft-Confidential" mark
\AddToShipoutPicture{%
    \AtPageCenter{%
        \makebox[0pt]{%
            \rotatebox[origin=c]{45}{%
                \scalebox{6}{%
                    \textcolor{gray!20}{DRAFT - CONFIDENTIAL}
                }
            }
        }
    }
}
\end{comment}

\title{LDPC-cat codes for low-overhead quantum computing in 2D}
\author{Diego Ruiz}
\email{diego.ruiz@alice-bob.com}%
\affiliation{Alice \& Bob, 49 Bd du Général Martial Valin, 75015 Paris, France}
\affiliation{Laboratoire de Physique de l'École Normale Supérieure, École Normale Supérieure, Centre Automatique et Systèmes, Mines Paris, Université PSL, Sorbonne Université, CNRS, Inria, 75005 Paris}
\author{Jérémie Guillaud}%
\affiliation{Alice \& Bob, 49 Bd du Général Martial Valin, 75015 Paris, France}
\author{Anthony Leverrier}%
\affiliation{Inria Paris, France}
\author{Mazyar Mirrahimi}%
\affiliation{Laboratoire de Physique de l'École Normale Supérieure, École Normale Supérieure, Centre Automatique et Systèmes, Mines Paris, Université PSL, Sorbonne Université, CNRS, Inria, 75005 Paris}
\author{Christophe Vuillot}%
\affiliation{Université de Lorraine, CNRS, Inria, LORIA, F-54000 Nancy, France}

\date{\today}

\begin{abstract}
Quantum low-density parity-check (qLDPC) codes are a promising construction for drastically reducing the overhead of fault-tolerant quantum computing (FTQC) architectures.
However, all of the known hardware implementations of these codes require advanced technologies, such as long-range qubit connectivity, high-weight stabilizers, or multi-layered chip layouts.
An alternative approach to reduce the hardware overhead of fault-tolerance is to use bosonic cat qubits where bit-flip errors are exponentially suppressed by design. 
In this work, we combine both approaches and propose an architecture based on cat qubits concatenated in classical LDPC codes correcting for phase-flips.
We find that employing such phase-flip LDPC codes provides two major advantages.
First, the hardware implementation of the code can be realised using short-range qubit interactions in 2D and low-weight stabilizers, which makes it readily compatible with current superconducting circuit technologies.
Second, we demonstrate how to implement a fault-tolerant universal set of logical gates with a second layer of cat qubits while maintaining the local connectivity.
We conduct a numerical brute force optimisation of these classical codes to find the ones with the best encoding rate for algorithmically relevant code distances. We discover that some of the best codes benefit from a cellular automaton structure. This allows us to define  families of  codes with high encoding rates and distances.
Finally, we numerically assess the performance of our codes under circuit-level noise.
Assuming a physical phase-flip error probability  $\epsilon \approx 0.1\%$, our $[165+8\ell, 34+2\ell, 22]$ code family allows to encode $100$ logical qubits with a total logical error probability (including both logical phase-flip and bit-flip) per cycle and per logical qubit $\epsilon_L \leq 10^{-8}$ on a $758$ cat qubit chip.
\end{abstract}


\maketitle

%%%%%%%%%%%%%%%%%%%%%%%%%%%%%%%%
% MAIN TEXT
%%%%%%%%%%%%%%%%%%%%%%%%%%%%%%%%

\section{\label{sec:introduction}Introduction}

\begin{table*}[t!]  % Use [t] to place the table at the top of the page
\centering
\begin{tabularx}{\textwidth}{@{} >{\centering\arraybackslash}m{0.23\linewidth} *{4}{>{\centering\arraybackslash}m{0.1875\linewidth}} @{}}
\toprule
\textbf{}                                                       & \begin{tabular}[c]{@{}c@{}}Surface code \\ + sc qubits~\cite{Fowler2012,mcewen2023relaxing} \end{tabular} & \begin{tabular}[c]{@{}c@{}}High-rate qLDPC \\  codes + sc qubits~\cite{bravyi2023}\end{tabular} & \begin{tabular}[c]{@{}c@{}}Repetition code \\ + cat qubits~\cite{Gouzien2023}\end{tabular} & \textbf{\begin{tabular}[c]{@{}c@{}}High-rate LDPC \\ code + cat qubits\end{tabular}} \\ \midrule \midrule
Short-range interactions                                                        & yes (2D)                                                                 & no (2D)                                                                    & yes (1D)                                                                     & yes (2D)                                                                               \\
Tanner graph degree                                             & 3-4                                                                   & 6                                                                       & 2                                                                       & 4                                                                                 \\ \midrule
$N_L=100$ footprints                                                       &                                                                  &                                                                      &                                                                       &                                                                                 \\
\multicolumn{1}{@{}l@{}}{\multirow{2}{*}{$\left.\begin{array}{@{}l@{}}\epsilon = 10^{-3} \\ \kappa_1/\kappa_2 = 10^{-4}\end{array}\right\}$ $\rightarrow \epsilon_L \leq 10^{-8}$}} & N = 33,700                                                          & 2,400 (N/14)                                                                    & -                                                                    & -                                                                              \\
 & -                                                          & -                                                                    & 2,100 (N/16)                                                                    & 758 (N/44)                                                                              \\
 \bottomrule                                                                             
\end{tabularx}
\caption{Technological cost versus footprint reduction trade-offs for recently proposed architectures, compared to the surface code. The technological cost includes the locality of qubit interactions and the degree of the Tanner graph (corresponding to the weight of the stabilizers), which sets the degree of qubit connectivity. The footprint corresponds to the total number of physical qubits $N$ required to implement a processor with $N_L = 100$ logical qubits, with a logical error rate per code cycle and per logical qubit of $\epsilon_L \leq 10^{-8}$, assuming circuit-level noise with a physical error rate $\epsilon = 10^{-3}$ for generic superconducting (sc) qubits or a ratio $\kappa_1/\kappa_2 = 10^{-4}$ for cat qubits, where $1/\kappa_1$ is the single-photon lifetime of the resonator hosting the cat qubit and $\kappa_2$ the two-photon stabilisation rate of the cat qubit, which ratio determines the error models of all cat qubit operations. Concretely, the value $\kappa_1/\kappa_2 = 10^{-4}$ corresponds to a circuit-level error model with state preparation and measurement infidelities $\epsilon_{\text{SPAM}} = 1.1\times10^{-3}$, CNOT gate infidelity of $\epsilon_{\text{CNOT}} = 1.6\times10^{-2}$ and idling errors of $\epsilon_{\text{idling}} = 1.1\times10^{-3}$, that is, all operations are noisier than for a depolarizing error model with strength $\epsilon = 10^{-3}$. We detail how these numbers were computed and provide a background discussion on cat qubits error models in Appendix~\ref{sec:archi_comp}.}
\label{tab:architectures_comparison}
\end{table*}

The discovery of quantum algorithms capable of solving certain computational problems exponentially faster than their classical counterparts has ignited a race in the development of quantum computers~\cite{Simon97,Shor1997}.  However, despite two decades of remarkable technological progress in building these machines, the quantum processors available today still lack the capability to solve problems of practical interest for research or business.
The primary challenge lies in the typical number of gates of quantum algorithms designed to address practical problems, reaching up to $10^7-10^{11}$ quantum gates for certain applications~\cite{beverland2022assessing,Daley2022,Hoefler2023,AWS_AlgorithmsSurvey}. Current processors, constrained by quantum decoherence to error rates around $10^{-3}-10^{-4}$ even for the best among them~\cite{moses2023,google2023,kim2023}, cannot reliably execute these algorithms.

The theory of quantum error correction and fault-tolerant quantum computing~\cite{Shor1995,Shor1996,Steane1996,Knill1998}, in principle, offers a solution to this problem. The ``algorithmic'', or logical, quantum information is encoded non-locally using error-correcting codes, implemented with a collection of physical quantum systems. When these codes are operated below their fault-tolerance threshold~\cite{Knill1998}, logical qubit errors can be arbitrarily suppressed by increasing the code distance $d$ (the minimum number of physical qubits that need to be acted upon to transition from a logical codeword to another), \textit{i.e.}\ by increasing the number of encoding physical qubits.
In practice, the implementation of error correction incurs a significant hardware overhead. For instance, it is the case for the surface code~\cite{Dennis2002,Kitaev2003,Fowler2012} —a particularly popular code due to its feasibility on a 2D grid of physical qubits using only nearest-neighbor interactions and displaying a notably high threshold— which was recently implemented on superconducting qubits to successfully demonstrate the suppression of logical errors with the code distance~\cite{google2023}. Despite the favorable attributes of this code, it is estimated that algorithms requiring on the order of 1,000 to 10,000 algorithmic qubits will demand several million physical qubits when accounting for the overhead of the fault-tolerant implementation with the surface code~\cite{Gidney2021, beverland2022assessing}. This significant overhead is partially explained by the relatively poor encoding rate of the surface code. For a $\llbracket n, k, d \rrbracket$ quantum code encoding $k$ logical qubits in $n$ physical qubits with distance $d$, the encoding rate is defined as the ratio of the number of logical qubits to physical qubits $k/n$. The encoding rate of the surface code is given by $1/n$. Given that the surface code distance $d$ scales as $\sqrt n$, this translates into exceedingly low encoding rates, approximately ranging from $1\%$ to $0.1\%$, for the practical distances of interest falling within the range of $10-30$. To reduce this overhead, three complementary approaches may be considered: (i) improve the quality of physical components to decrease the required code distances, thereby improving encoding rates, (ii) develop new error-correcting codes with superior performance —higher thresholds or better encoding rates—, (iii) optimize the synergy between the code and its physical component noise structure.

An important body of theoretical and experimental research aims at addressing the approach (i),  either by better isolating the physical qubits from their noisy environment or by designing and realizing qubits that are naturally protected against the dominant noise sources by virtue of some symmetries~\cite{brooks2013protected}.

The main directions pursued in (ii) are to find codes with a sufficiently large distance for the targeted applications, with an increased threshold to alleviate constraints on hardware quality, with better logical error scaling, or to increase the encoding rate to reduce the footprint of the architecture. The class of quantum low-density parity-check (qLDPC) codes has been the focus of recent intensive research~\cite{PRXQuantum.2.040101}, leading to a series of theoretical breakthroughs, notably the discovery of ``good'' qLDPC codes~\cite{panteleev2022,leverrier2023,DHL23} with a non-vanishing encoding rate and a distance growing linearly with the block size $n$, along with the construction of efficient decoders for these codes~\cite{10103665,gu2023single}. These theoretical advances have inspired the systematic search and the discovery of small quantum LDPC codes with higher encoding rates than the surface code~\cite{bravyi2023}, or code concatenation between LDPC codes and surface codes in order to improve the overall rate~\cite{pattison2023hierarchical,gidney2023yoked}. However, implementing these superior codes incurs a higher technological cost than the surface code for a fundamental reason~\cite{Baspin_2022}. Indeed, Bravyi \textit{et al.} showed that the performance of a \textit{local} quantum error correcting code on a 2D lattice is upper bounded by $kd^2 = O(n)$~\cite{bravyi2010}. Because the surface code saturates this bound, improving upon it (with a 2D architecture) necessarily requires \textit{non-local} interactions -- \textit{i.e.}\ long-range interactions -- in the processor. While this property is feasible for certain physical platforms like neutral atoms~\cite{bluvstein2023logical,xu2023}, it is more challenging to realize with superconducting circuits, although the biplanarity property of some qLDPC codes may help~\cite{bravyi2023}. Finally, the fault-tolerant construction of a universal set of logical gates which can be reasonably parallelized on these qLDPC codes is still an active research problem ~\cite{gottesman2013,breuckmann2017,krishna2021,cohen2022,jochym2019fault,quintavalle2023partitioning,breuckmann2022fold,cowtan2023,zhu2023non,barkeshli2023higher,lavasani2019universal}.

Regarding the synergy between the error-correcting code and its physical component noise structure (iii), recent proposals suggest concatenating bosonic qubits with carefully selected properties into error-correcting codes tailored to suit these qubits' characteristics~\cite{fukuiHighThresholdFaultTolerantQuantum2018, vuillotQuantumErrorCorrection2019,Noh2020,Joshi2020,Guillaud2019,kubica2022erasure}. In particular, dissipatively stabilized cat qubits~\cite{Leghtas2015, Mirrahimi2014} stand out among bosonic qubits due to their remarkable property that the bit-flip error rate is exponentially suppressed with the average number of photons in the cat qubit, at the cost of a linear increase in the phase-flip error rate. This drastic scaling difference between these two types of errors allows for the realization of a qubit with extremely biased noise even for modest numbers of photons where bit-flip errors are many orders of magnitude rarer than phase-flip errors $\eta\ \dot{=}\ p_Z/(p_Y+p_X) \gtrsim 10^8$, as demonstrated experimentally~\cite{Lescanne2020,Berdou2022,reglade2023,marquet2023}. Such macroscopic bit-flip lifetimes may enable to build logical qubits with applications-relevant logical error rates by correcting only phase-flip errors, thereby reducing the complexity of quantum error correction to that of classical error correction. The simplest way to achieve this is to concatenate cat qubits in a repetition code protecting against phase-flip errors, an architecture that has been the focus of many recent theoretical proposals and optimizations~\cite{Guillaud2019,Guillaud2020,Chamberland2022,Gouzien2023,Xu2023squeezed,LeRegent2023highperformance}. The stabilizers of the (distance-$d$) phase-flip repetition code are $\{X_i X_{i+1}\}_{i \in \{0,...,d-2\}}$ and the logical Pauli operators are $X_L = X_0$, $Z_L = \bigotimes_{i=0}^{d-1} Z_i$ and $Y_L = -i Z_L X_L$. The operator $X_L$ has weight one, as expected, because the code does not possess bit-flip error-correcting capabilities, and protection against bit-flip errors is entirely achieved at the level of the cat qubit. Also, observe that this architecture does not have a conventional threshold, as the physical phase-flip rate of cat qubits increases linearly with the average photon number, eventually surpassing the phase-flip threshold of the repetition code (this fact is not limiting in practice, as discussed in Appendix~\ref{sec:archi_comp}). Despite this fact, the passive suppression of bit-flip errors at the level of the low hardware layers and the use of the error-correcting code for phase-flip correction exclusively enables to reach sufficiently low logical error rates for practical applications with a competitive overhead compared to other approaches~\cite{Chamberland2022,Gouzien2023}. In the regime of moderate noise bias $\eta \approx 10^2$ where active correction against bit-flips at the level of the code remains necessary, it was recently shown how qLDPC codes could be bias-tailored to improve their performance~\cite{Roffe2023}.

The main result of our work is to further reduce the footprint of the cat qubit architecture by replacing the repetition code with classical LDPC codes featuring improved encoding rates, while retaining the capability of executing a universal set of fault-tolerant logical gates, at the cost of a minimal technological overhead. The central idea we exploit is that it is possible to construct 2D codes that outperform the 1D repetition code without sacrificing locality. Indeed, the ``classical version'' of the BPT bound~\cite{bravyi2010} is $k\sqrt{d} = O(n)$ for 2D codes, and repetition codes ($k=1, d=n$) are far from saturating this bound. Our numerical search for \textit{local} LDPC codes in 2D with favorable parameters allows us to identify codes with the desired distances $\approx 10-30$ that exhibit up to $5\times$ higher encoding rates than the repetition code, at the cost of increasing the weight of stabilizers from two to four. Furthermore, we show how the lattice surgery schemes of the repetition code architecture can be adapted to these codes in order to perform a universal set of fault-tolerant logical gates. More precisely, we propose to use an ancillary layer of logical qubits encoded in repetition codes in order to realize logical entangling gates between logical qubits encoded in the phase-flip LDPC codes and to inject magic states.

We provide in \autoref{tab:architectures_comparison} a comparison of the footprint and the key technological assumptions of our architecture alongside some of the best alternatives to build a 100-logical qubit quantum processor with a logical error rate $\epsilon_L \leq 10^{-8}$ using superconducting circuits. This illustrative target was chosen for two reasons. First, these numbers roughly correspond to the entrance into the ``fault-tolerant'' regime where quantum hardware becomes able to solve useful computational problems that are beyond the reach of classical methods. 
For instance, some open problems in quantum materials could be tackled with 100 logical qubits and a hybrid-classical approach~\cite{Bauer2016}, the simulation of the dynamics of a 2D Hubbard model with $10\times10$ lattice sites and 100 hundred particles with $10^6$ logical gates could be performed with 200 logical qubits~\cite{daley2022practical}, and the simulation of 3D spinful jellium with $1.8\times10^7$ T gates could be done with only 123 logical qubits~\cite{babbush2018encoding}.
Slightly increasing these numbers gives access to more applications, \textit{e.g.}\ the quantum phase estimation of FeMoCo active spaces to within chemical accuracy using a 2,142-logical qubit quantum computer with $5.3\times10^{9}$ Toffoli gates~\cite{Lee2021}. Second, the reported footprint of our architecture corresponds to the case where bit-flip errors are entirely suppressed using the passive error correction provided by cat qubits, such that the active error correction code is devoted exclusively to phase-flip error correction. The $\epsilon_L \leq 10^{-8}$ target corresponds to a physical bit-flip time at the level of the cat qubit of $T_X \sim 13$ minutes, which we deem to be reasonably close to the bit-flip times already demonstrated experimentally~\cite{Berdou2022, marquet2023, reglade2023}.

The remainder of the article is structured as follows. Section~\ref{sec:codessearch} presents the method and results of our numerical exploration for high-rate phase-flip codes that are local in two dimensions. In Section~\ref{sec:numerics}, we numerically assess the performance of the best codes we found under phenomenological and circuit-level error models. Subsequently, in Section~\ref{sec:gates}, we demonstrate how a universal set of fault-tolerant logical gates can be implemented on these codes using ancillary logical qubits encoded in repetition codes. Remarkably, this construction can be implemented using 
the same hardware as for operating the code \textit{i.e.} with local, short-range connectivity. Finally, the experimental implementation of our architecture in a superconducting circuits platform is discussed in Section~\ref{sec:Experiment}, where we show that it can be realized using currently available technological building blocks.

\section{\label{sec:codessearch}High-rate local phase-flip codes}

%-------------------------------%
\begin{figure}[t!]
    \centering
    \includegraphics[width=\columnwidth]{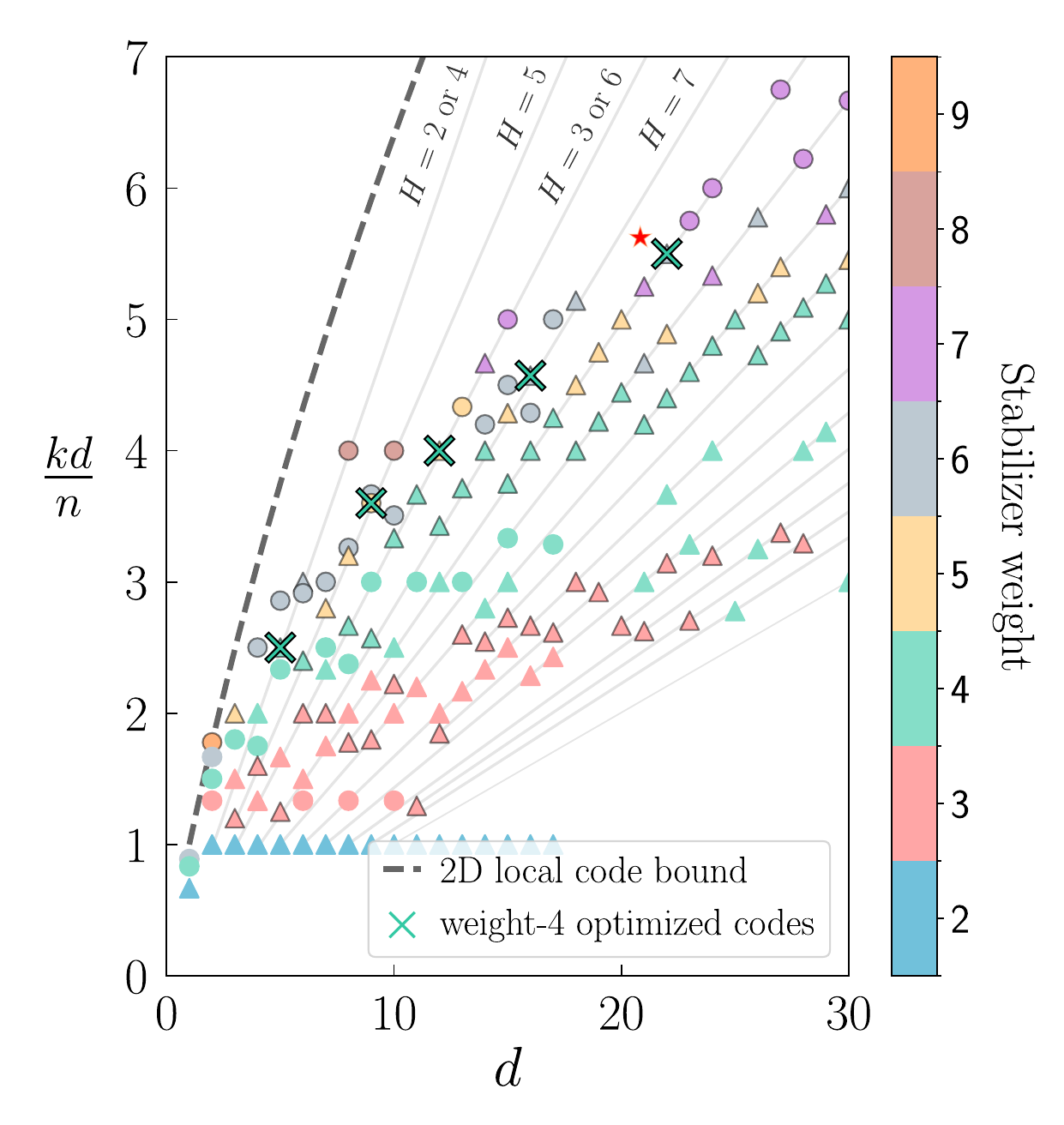}
    \vspace{-0.5cm}
    \caption{\label{fig:AllCodes}
    Overhead reduction factor over the repetition code $kd/n$ of 2D local phase-flip codes, as a function of the code distance. Each marker represents a single code and the color indicates the stabilizer weight. These codes have local stabilizers fitting within a $3\times3$ grid of data qubits, which are invariant under vertical and horizontal translations on the grid of qubits. The triangles highlight the family of cellular automaton codes, which feature attractive properties for the quantum architecture (see main text). All codes are constructed on lattices of size $H\times L \leq 17\times 17$, with periodic boundary conditions on the lateral sides --added for simplicity, but codes with essentially identical parameters exist without these conditions (Section~\ref{sec:Experiment})-- and only the best codes are displayed.
    The codes identified by a cross have been found by allowing the shape of the (weight-4) stabilizers to differ at each row of the lattice. The 2D code performance upper bound is also indicated~\cite{bravyi2010}. 
    Markers with a contour indicate stabilizer shapes which span three rows. Finally, we display grey lines with slopes $(m-1)d/H$, characteristic of cellular automaton codes (see main text).
    }
\end{figure}
%-------------------------------%

%-------------------------------%
\begin{figure*}[t!]
    \centering
    \includegraphics[width=0.9\textwidth]{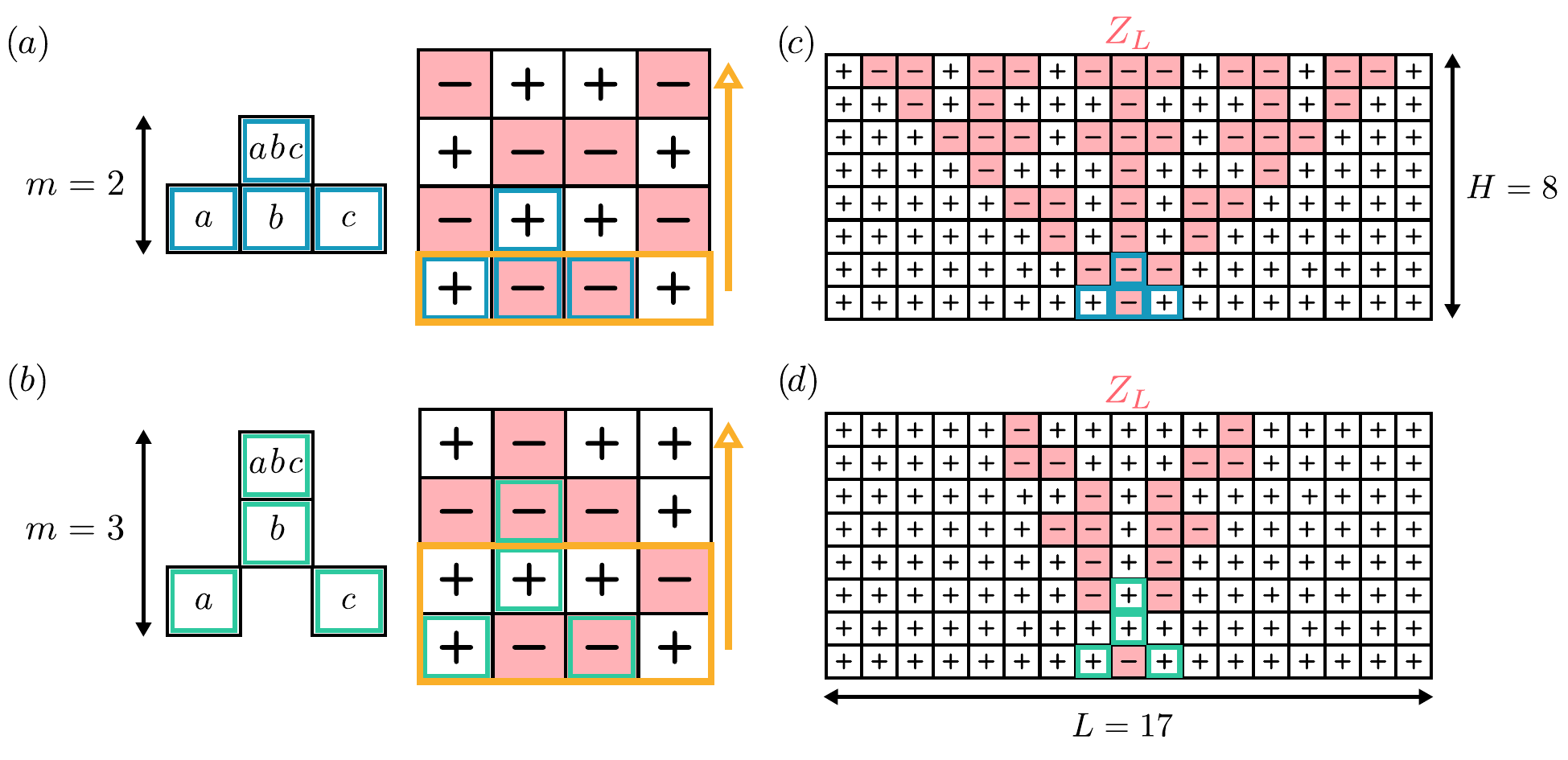}
    \vspace{-0.5cm}
    \caption{\label{fig:automate}
    Representation of phase-flip cellular automaton codes on a $H \times L$ lattice of qubits. Cellular automaton codes are a family of local codes characterized by ``pointed''-shape of stabilizers, \textit{i.e.}\ stabilizers acting on a single qubit of the top row of their support. As a consequence, a stabilizer shape spanning $m$ rows corresponds to a code of dimension $k = (m-1)L$, and the $2^k$ codewords are uniquely determined by the $|\pm\rangle$ state of the bottom $(m-1)$ rows of qubits. The entire codewords can be constructed row by row by successive applications of the cellular automaton rule corresponding to the stabilizer, as depicted (yellow arrow) for (a) a ($m=2, L=4$) code encoding $4$ logical qubits and (b) a ($m=3, L=4$) code encoding $8$ logical qubits. 
    In (c-d), we represent a $Z_L$ logical operator of minimal weight for the two codes corresponding to the stabilizer shapes in (a-b), respectively. Note that any horizontal shift of this logical operator remains a valid logical operator, thus defining a basis of minimum-weight logical operators. The ``fractal'' support of the logical operator justifies the occasional reference to cellular automaton codes as fractal codes.
    Note that the code (d) has a smaller distance than the code (c). This is however compensated by a doubled encoding rate, such that the overal overhead reduction factor is larger for the same distance.
     }
\end{figure*}
%-------------------------------%

%-------------------------------%
\begin{figure*}[t!]
    \centering
    \includegraphics[width=0.9\textwidth]{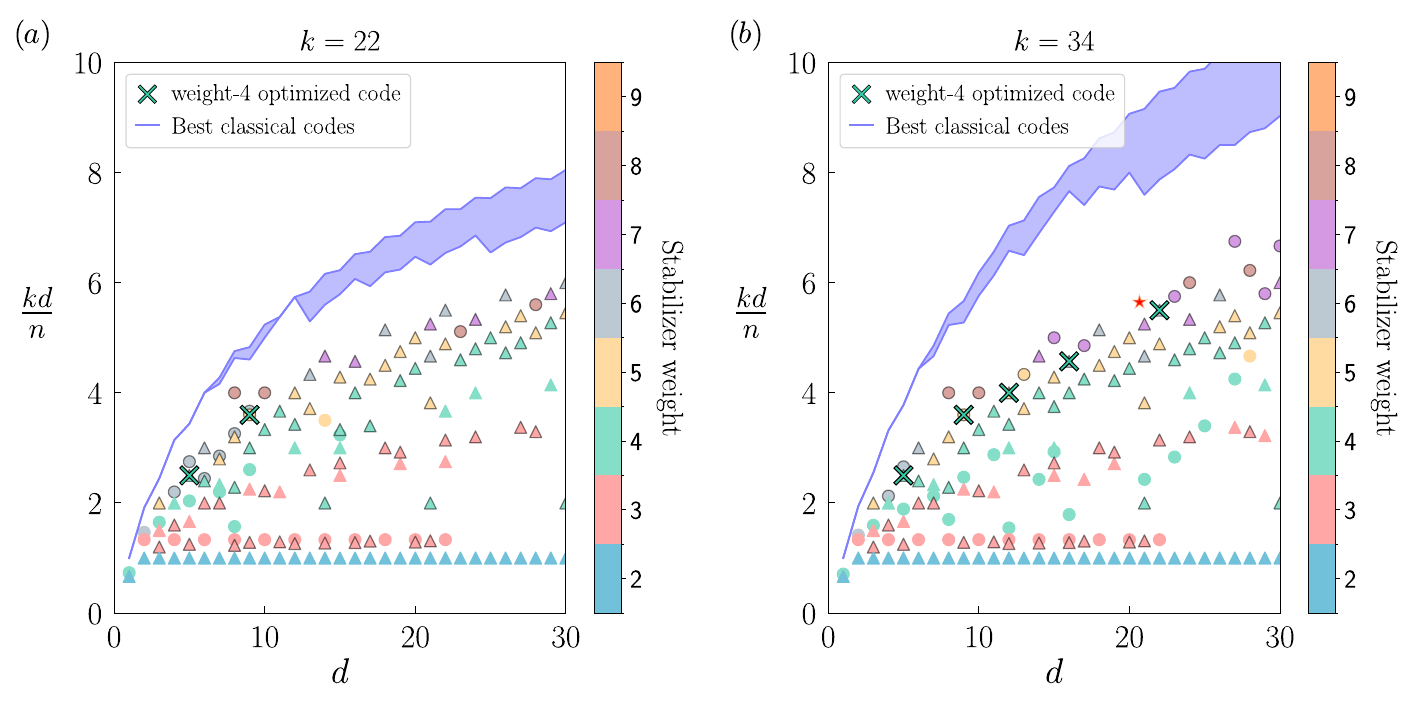}
    \vspace{-0.5cm}
    \caption{\label{fig:fixed_k}
    Performance of quasi-cyclic 2D local codes encoding (a) $k=22$ logical qubits and (b) $k=34$ logical qubits as a function of the code distance.
    Each triangle and circle marker represents a code with a single stabilizer shape fitting within a 3 by 3 square.
    The other stabilizers are obtained by translating this shape horizontally and vertically. The codes are constructed on lattices of size up to $L = H = 34$, with periodic boundary conditions on the lateral sides.
    Optimized codes with possibly different weight-4 stabilizers on each row are identified by a cross.
    To compare the performance of our local codes in an absolute sense, we represent, for a fixed number of logical qubit, the theoretical upper bound for the overhead reduction factor together with the best known classical codes~\cite{Grassl:codetables} (thick blue curve, where the top part indicates the best possible code and the bottom part the best known code). Quite remarkably, for the parameters $k$ and $d$ we consider, the local codes we found are close to the best (potentially non-local) existing classical codes. 
    }
\end{figure*}
%-------------------------------%

In this section, we turn our attention to the search for small phase-flip LDPC codes that have higher encoding rates than repetition codes at identical distances. Classical linear codes are usually expressed as bit-flip codes, with codewords represented as $\{0, 1\}$ bit-strings. Here, to comply with the literature on cat qubits, we work in the dual basis to construct phase-flip codes, for which the stabilizer group $\mathcal{S}$ is a strict subgroup of the Pauli group with $X$-type stabilizers exclusively $\mathcal{S} \subseteq \{I, X\}^{\otimes n}$. As a consequence, there exists a basis of the codespace containing exclusively separable states of the form $|\{+,-\} \rangle ^{\otimes n}$. It is noteworthy that concatenating cat qubits in these codes yields an architecture that, like the repetition code, lacks the capability to correct bit-flips (logical $X_L$ operators can have weight one) and does not have a threshold. Similarly to the repetition code, we will simply refer to the phase-flip distance $d_Z$ of these codes as the distance $d$ with the implicit understanding that the distance against bit-flips is $d_X=1$.

We consider codes defined on a 2D rectangular lattice of qubits, with width $L$ and height $H$. As the goal is to improve the encoding rate compared to the repetition code, the metric that will be important for comparison is $kd/n$. This stems from the fact that for a repetition code, $k=1$ and $d=n$, such that $k$ logical qubits encoded in $k$ repetition codes correspond to $kd/n = 1$. Therefore, the value of $kd/n$ for other phase-flip codes can be directly interpreted as the ``overhead reduction factor'' (over repetition codes) of the number of physical qubits to implement a given number of logical qubits with a specific distance. Note that codes that have the same distance do not necessarily have identical logical error rates, as other code-dependent parameters come into play, such as the threshold, stabilizer weights, or the depth of stabilizer measurement circuits. This ratio is nonetheless a good proxy for the performance of the codes, as confirmed numerically in \ref{sec:numerics}, where we analyze in more details the best of these codes using circuit-level noise simulations.

We conduct the search for codes under two ``tunable'' technological constraints, allowing for systematic performance evaluation starting from the technologically simplest codes and progressing towards more complex ones. This approach will later enable us to settle on a performance versus technological cost trade-off that seems reasonable. Specifically, these two constraints are:
\begin{itemize}
    \item \textit{Short-range interactions.} Knowing that theoretical improvements can be achieved compared to the repetition code while retaining locality in 2D, we impose this constraint and consider nearest neighbor stabilizers, and subsequently explore next-nearest neighbor stabilizers. This is in contrast with previous approaches~\cite{bravyi2023,xu2023} that required long-range interactions.
    \item \textit{Connectivity degree.} From a hardware perspective, low connectivity degrees are desirable as they simplify numerous technical issues such as frequency crowding, cross-talk problems, provide more margin for chip fabrication variability, and simplify the design of the microwave filtering part, among other advantages. In our simplified model, the connectivity degree of the processor is simply determined by the weight of the stabilizers, so that we focus on low-weight stabilizers (which usually yield higher thresholds, when entangling gates of the stabilizer measurement circuit are applied sequentially).
\end{itemize}

The efficient search for codes is carried out as follows. \\

\textit{Unique stabilizer shape.} We first consider codes where all the stabilizers are generated by horizontal and vertical translations of a single generator stabilizer on the 2D lattice, so that the ``shape'' of the stabilizer is unique. We fix the dimensions $(H, L)$ of the lattice, thereby determining the number of physical data qubits $n = HL$, and consider periodic boundary conditions on the lateral edges, which allows us to avoid specifying the precise shape of the code lateral boundaries. In Section~\ref{sec:Experiment}, we show how to remove the periodic boundary conditions without significantly affecting the code parameters. This is crucial as we are ultimately interested in an implementation which avoids having to connect physical qubits on opposite edges. Code locality is enforced by restricting the search to stabilizers that fit within a $3\times3$ grid of neighboring data qubits. We systematically test all possible stabilizer shapes among the 512 possibilities $\{I, X\}^{\otimes{9}}$. By construction, the stabilizers have weight at most 9, and at most next-nearest neighbor physical extension. Each possible stabilizer shape yields a code for which we calculate the dimension $k$ and distance $d$. Since these codes are linear, one can put their parity-check matrix (defined by the stabilizers) into normal form 
%a systematic approach is to transform the parity-check matrix (defined by the stabilizers) into normal form 
to obtain a basis of $k$ codewords of the codespace. The distance of the code may then be computed exactly by constructing the $2^k$ codewords, as the code distance corresponds to the minimal non-zero Hamming weight of the codewords. An alternative, sometimes more efficient method for calculating the distance is to map this problem to a satisfiability (SAT) problem to leverage efficient available SAT solvers~\cite{z3}, as we detail in Appendix~\ref{subsec:SAT_distance}. 

The results are depicted in Figure~\ref{fig:AllCodes}, where we show the best codes we found. Each data point corresponds to a code associated with a stabilizer shape, and we compare the overhead reduction factor $kd/n$ as a function of the code distance $d$. As expected, codes with high stabilizer weights generally exhibit superior performance, and codes with weight-2 stabilizers correspond to repetition codes ($kd/n = 1$). As the distance increases, the overhead reduction factor of the codes we discovered deviates from the BPT bound~\cite{bravyi2010}, and our most efficient codes appear to follow a power law slightly smaller than a square root as a function of the code distance. Note that families of codes saturating the bound are known to exist, but these constructions either involve high-weight stabilizers that are ruled out by our technological constraints~\cite{yoshida2013}, or yield codes with relatively low number of logical qubits $k$ but extremely large distances $d$~\cite{tan2023} which is not our regime of interest.

\textit{Cellular automaton codes.} Interestingly, many of the best codes we found belong to the family of cellular automaton codes~\cite{newman1999glassy,bravyi2010,chowdhury1994,san2023cellular} (also referred to as ``fractal'' or ``Fibonacci'' codes in the literature~\cite{yoshida2013,nixon2021}), which is a strict subset of the entire exploration space. These codes were already studied for their better asymptotic scaling compared to the repetition code~\cite{bravyi2010,yoshida2013}. Additionally, a decoder was constructed in~\cite{nixon2021} (assuming perfect syndrome measurements) and demonstrated that these code are robust against string-like errors across the lattice. These codes, identified by triangles in Figure~\ref{fig:AllCodes}, are characterized by the fact that the stabilizer has a ``pointed'' shape, \textit{i.e.}\ the stabilizer non-trivially acts on only one of the qubits in the top row of its support. These codes have been known to achieve a good performance, due to the ``fractal'' nature of the support of logical operators, which yields large distances compactly~\cite{bravyi2010,yoshida2013,nixon2021}. 

Cellular automaton codes have the property that codewords can be systematically constructed using a cellular automaton rule corresponding to the stabilizer's shape. This is illustrated in Figure~\ref{fig:automate}(a-b): a cellular automaton code with the stabilizer shape extending over $m$ rows ($m=2$ or 3 here) encodes $k=(m-1)L$ logical qubits with a rate $(m-1)/H$. Indeed, each of the $2^k$ codewords is entirely determined by the values of the physical qubits in the lowest $m-1$ rows by applying the cellular automaton rule successively to determine the value of all physical qubits. Consequently, the codewords can be interpreted as the vertical evolution of a 1D cellular automaton, with time progressing upwards and the stabilizers dictating the evolution rule.

The minimum-weight logical $Z_L$ operators of the code can be defined as the product of $Z$ operators acting on the locations of $\ket{-}$ states in the codewords with the minimal non-zero Hamming weight. As an example, we depict in Figure~\ref{fig:automate}(c-d) a minimum-weight logical operator $Z_L$ (in red) of the codes defined by the stabilizers in (a) and (b), respectively. We numerically observe that, for a lattice with sufficient width (about twice its height) or when periodic boundary conditions are removed (see Section~\ref{sec:Experiment}), the codewords with the smallest Hamming weight, which determine the distance, are the ones with a single $\ket{-}$ state in the bottom row for most cellular automaton codes (as in Figure~\ref{fig:automate}(c-d)).

Since horizontal translations of logical operators on the lattice are also logical operators, a minimal-weight logical basis of the code can be formed easily. The $(m-1)L$ logical qubits of this basis consist of a single $\ket{-}$ state in the $m-1$ bottom rows and extend in a ``fractal'' manner to the rows above. Thereby, cellular automaton codes enable to  arbitrarily tune the number of logical qubits by changing the width of the lattice, maintaining a constant distance as long as the height remains unchanged. Alternatively, the distance can be improved by increasing the height of the lattice. The overhead reduction factor is then given by $(m-1)d/H$ forming the gray characteristic lines in Figure~\ref{fig:AllCodes}.

Note that with periodic boundary conditions, the codewords may ``loop'' to the opposite side of the code when the lattice width is too small compared to its height, potentially reducing the distance of the code.  However, this problem disappears when removing the periodic boundary conditions, as detailed in Section~\ref{sec:Experiment}.

We compare in Figure~\ref{fig:fixed_k} the (local) codes we found to the best possible classical codes (without any locality constraints) for fixed code dimensions (a) $k = 22$ and (b) $k=34$. The upper part of the curve represents the theoretical bound for the best codes, while the lower part represents the best-known classical codes~\cite{Grassl:codetables}. Remarkably, we observe that for a small number of logical qubits and relevant distances, the codes we found are close to the best possible codes. As the number of encoded logical qubits or the distance increase, our constructed codes deviate further from the best possible classical codes.

\begin{table*}[t!]
\centering
\begin{tblr}{
 colspec = {Q[c,m,10em]|Q[c,m,3em]|Q[c,m,7em]|Q[c,m,16em]},
   }
  \hline
  $[n,k,d]$ & $kd/n$ & $(H, L = L^* + \ell)$ & Stabilizer shapes (bottom to top) \\
  \hline
  $[20+4\ell,10+2\ell,5]$ & $2.5$ & (4, $5+\ell$) &  \begin{minipage}[c][2em][c]{10em}
    \includegraphics[width=4.75em]{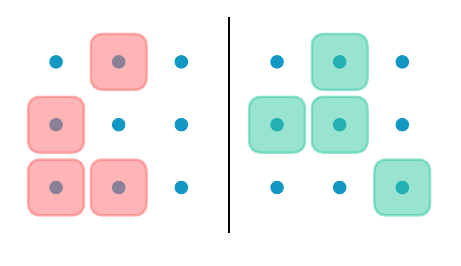}
  \end{minipage}  \\
  \hline
  $[55+5\ell,22+2\ell,9]$ & $3.6$ & (5, $11+\ell$) &  \begin{minipage}[c][2em][c]{10em}
    \includegraphics[width=6.75em]{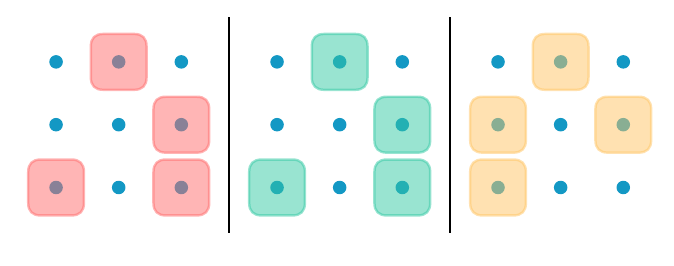}
  \end{minipage} \\
  \hline
  $[78+6\ell,26+2\ell,12]$ & $4$ & (6, $13+\ell$) &  \begin{minipage}[c][2em][c]{10em}
    \includegraphics[width=9em]{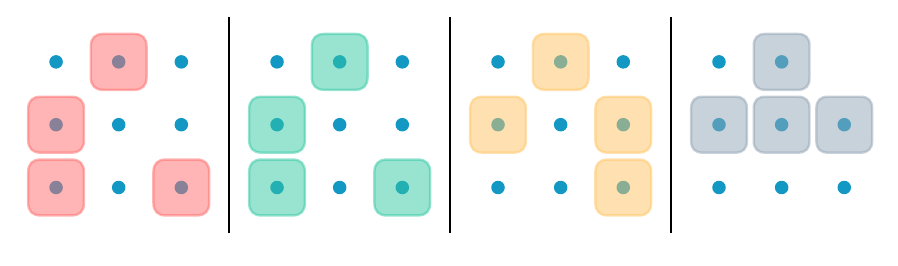}
  \end{minipage} \\
  \hline
  $[119+7\ell,34+2\ell,16]$ & $4.6$ & (7, $17+\ell$) & \begin{minipage}[c][2em][c]{10em}
    \hspace*{-0.5em}\includegraphics[width=11em]{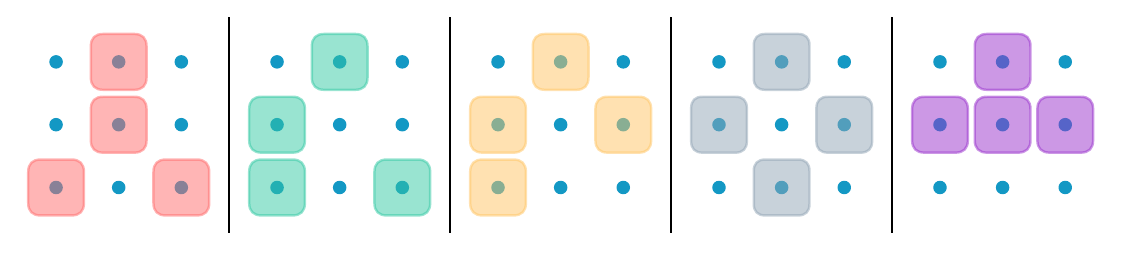}
  \end{minipage} \\
  \hline
  $\ [136+8\ell,34+2\ell,22]^{\etoile}$ & $5.5$ & (8, $17+\ell$) &  \begin{minipage}[c][2em][c]{10em}
    \hspace*{-1.5em}\includegraphics[width=13em]{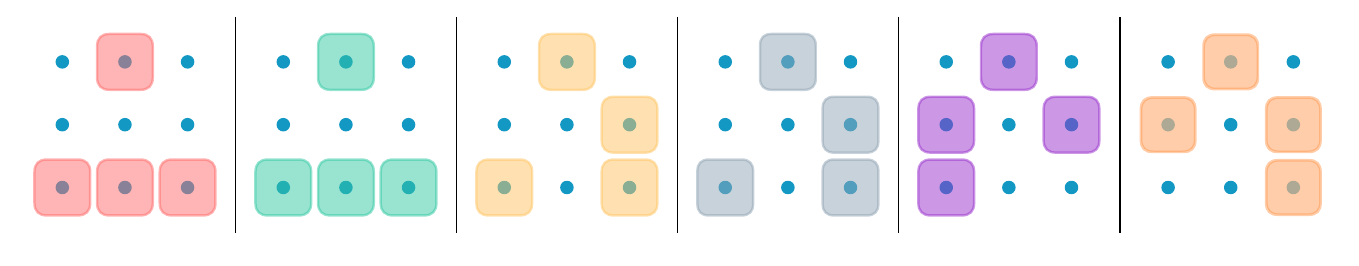}
  \end{minipage} \\
  \hline
\end{tblr}
\caption{Optimized cellular automaton codes on a $H\times L$ lattice of qubits with periodic boundary conditions on the lateral sides. These codes feature a different stabilizer shape for each row. $L^*$ corresponds to the minimal width $L$ required to achieve the full code distance set by the value of $H$: increasing values of $L \geq L^*$ ($H$ fixed) yields codes with the same distance $d$, but with code dimension $k = (m-1)L$ and code size $n = HL$, therefore, identical overhead reduction ratio $kd/n$. Here and throughout the paper, the red star identifies the $[136+8\ell,34+2\ell,22]$ code family. For the sake of completeness, we mention that in rare cases, interference phenomena reduce the distance for very specific values of $L > L^*$, as explained in the main text. These features disappear when the periodic boundary conditions are removed (see Section~\ref{sec:Experiment}).
}
\label{tab:codes}
\end{table*}

Note that, in the literature, cellular automaton codes correspond to the case $m = 2$, and the codes with $m\geq3$ are not strictly speaking cellular automaton codes.
For higher values of $m$, the code distance increases more slowly with $H$, as can be seen in Figure~\ref{fig:automate}(d) where the support of the logical operators $Z_L$ is smaller.
However, as these codes encode $m-1$ rows of logical qubits instead of one, the rate of the code is increased. As more logical qubits are encoded, the overall code parameters are improved, at the cost of stabilizers with a slightly longer range, as identified in Figure~\ref{fig:AllCodes} by markers with contour for codes where $m=3$.

\textit{Multiple stabilizer shapes.} We then expand the search space by considering codes not defined by a unique stabilizer shape. 
Ideally, we would like to explore the space where all stabilizer shapes and weights are different, but this represents too large a space for an exhaustive search. 
To reduce the size of the space, we choose to explore only cellular automaton codes, because the calculation of the code distance and dimension is simplified due to the specific structure of these codes. We further restrict the search to weight-4 stabilizers, which we deem to be the best compromise between stabilizer weight and the overhead reduction factor $kd/n$ from the results of the previous search (Figure~\ref{fig:AllCodes}). Finally, we restrict ourselves to cellular automaton codes for which all of the stabilizers in a given row of the lattice are identical (but can differ from one row to the other). These codes are quasi-cyclic, since parity checks are invariant under a cyclic permutation of the qubits of period $H$, when the qubits are numbered in column-major order~\cite{townsend1967}. Even with these restrictions, a naive approach to the search is not tractable, and a SAT solver was also used to compute the distance (see Appendix~\ref{subsec:SAT_shape}).

\begin{figure*}[t!]
    \centering
    \includegraphics[width=0.7\textwidth]{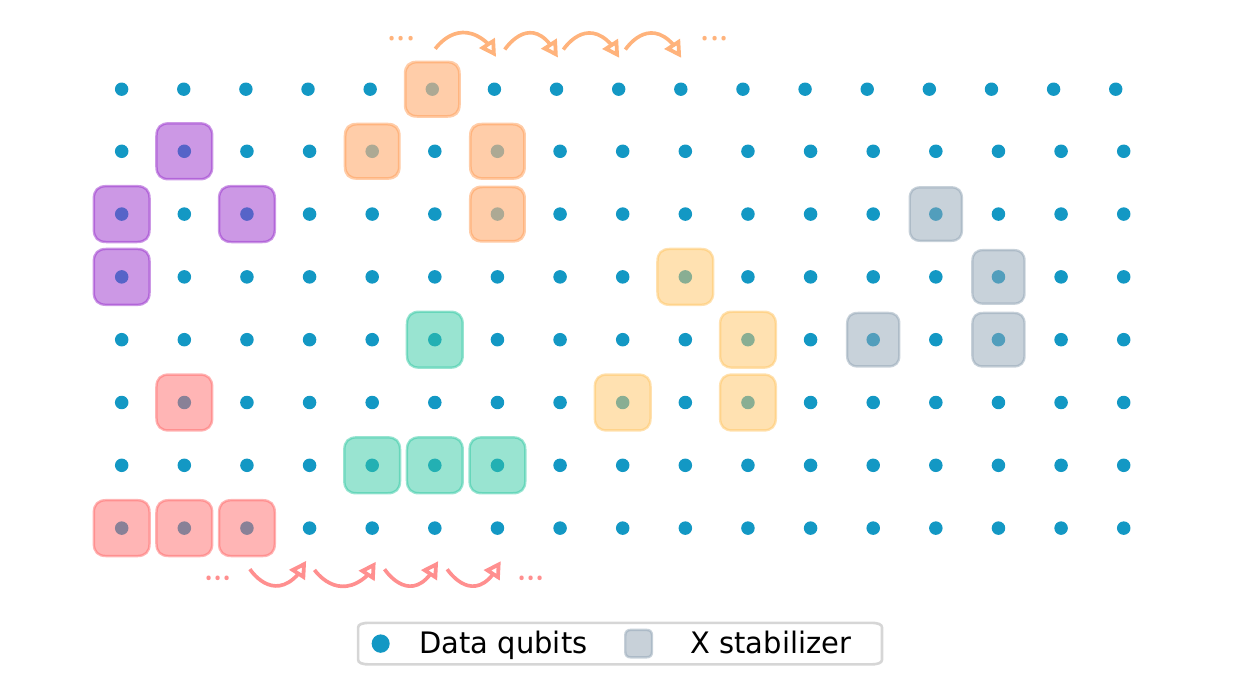}
    \vspace{0cm}
    \caption{\label{fig:Code}
    Layout of the $[136,34,22]^{\etoile}$ phase-flip code. The data qubits are represented as blue dots and the 6 patterns of $X$-type stabilizers as colored squares. The code belongs to the family of quasi-cyclic codes~\cite{townsend1967}, the weight-4 stabilizer on each row is repeated $L=17$ times in the horizontal direction (for a total of 85 stabilizers). Here, the code is represented with periodic boundary conditions on the lateral sides, but this constraint can be safely removed for an experimental realization (see Section~\ref{sec:Experiment}).
    }
\end{figure*}

To assess the gain in performance due to these optimizations, we add the new codes we find (marked by crosses) to Figures \ref{fig:AllCodes} and \ref{fig:fixed_k}. The specific structure of these codes can be found in Table~\ref{tab:codes}.
The optimization significantly improves the performance, as can be seen in Figure~\ref{fig:AllCodes}: the codes with multiple weight-4 stabilizer shapes achieve comparable performances as the codes with a single stabilizer shape of weights 5 to 7. We show the precise layout of one of the interesting codes we found, $[136, 34, 22]$ in Figure~\ref{fig:Code}. This optimized code reduces the overhead by a factor $kd/n = 5.5$ for distance $d=22$, nearly doubling the performance over the previously existing cellular automaton codes.

%-------------------------------%
%-------------------------------%

Finally, we searched for more codes by considering different stabilizer shapes on each site of the lattice. The space this represents being too large for an exhaustive search, we restricted ourselves to small lattices size ($H = 5$ and $L=10$) and looked for cellular automaton codes with $m=2$. On these small lattices, we did not find significant improvements over the codes we already found. We leave the search for such codes on larger lattice sizes to future work.

\section{\label{sec:numerics}Numerical estimation of the logical error probability}

%-------------------------------%
\afterpage{%
\begin{figure*}[!htbp]
    \centering
    \includegraphics[width=0.95\textwidth]{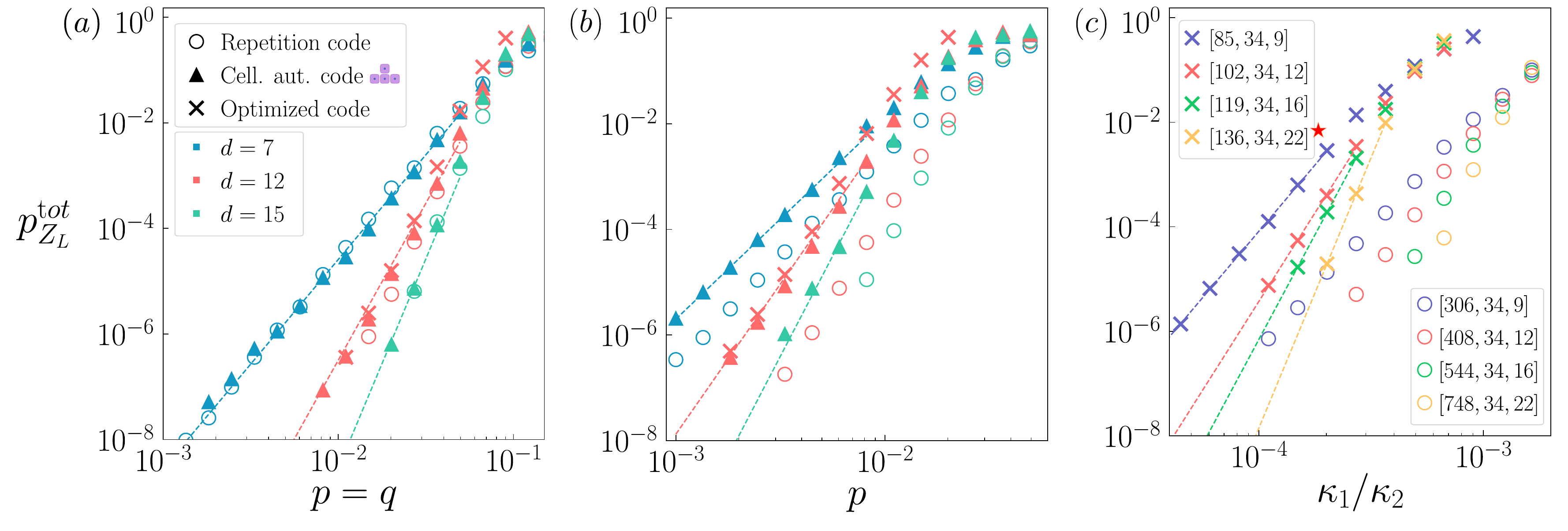}
    \vspace{-0.5cm}
    \caption{\label{fig:numerics}
    Logical phase-flip error probability per logical qubit after $d$ error correction cycles $p_{Z_L}^{\rm tot}$ as a function of the physical error probability, under (a) a phenomenological error model, (b) a generic phase-flip circuit level error model, or (c) a cat qubit circuit-level error model, which are detailed in Table~\ref{tab:error_models}. Repetition codes are identified by circles, $\tetris$ cellular automaton codes by triangles, and optimized codes from Table~\ref{tab:codes} by crosses.
    In (a)-(b), repetition codes are compared to the $\tetris$ cellular automaton code family to understand generic features of these codes. We observe that codes with identical distances have identical logical error probability under a phenomenological error model (a), which indicates that the repetition code and cellular automaton code have similar thresholds. In (b), the effect of the depth of stabilizer measurement circuits becomes visible: the threshold of weight-4 codes (cellular automaton and optimized codes) is roughly divided by a factor two compared to the weight-2 repetition codes. In (c), we compare the performance of the best codes we found (Table~\ref{tab:codes}) under circuit-level noise, since these codes are the ones that will be implemented in practice. The fit in (c) is used to extrapolate the logical error rate of the $d=22$ code to $\kappa_1/\kappa_2=10^{-4}$.
    }
\end{figure*}
%-------------------------------%
\begin{table*}[!htbp]
\small % Reduce font size
\newcolumntype{C}{>{\centering\arraybackslash}X}
\begin{tabularx}{0.9\textwidth}{@{} >{\hsize=0.4\hsize}C >{\hsize=0.5\hsize}C >{\hsize=1.1\hsize}C >{\hsize=1.3\hsize}C >{\hsize=1.4\hsize}C @{}} 
\toprule
                                           &               & (a) Phenomenological & (b) Generic circuit-level & (c) Cat qubit circuit-level                                         \\ 
\midrule
\midrule
\multirow{2}{*}{$\mathcal{P}_{|+\rangle}$} & $Z$           & -                    & $p$                       & $\bar{n}\kappa_1T_{\text{prep}}$                                         \\
                                           & Idle          & $p$                  & $p$                       & $\bar{n}\kappa_1T_{\text{prep}}$                                         \\ 
\midrule
\multirow{2}{*}{$\mathcal{M}_{X}$}         & $1 - \mathcal{F}$ & $p=q$                  & $p$                       & $\bar{n}\kappa_1T_{\text{meas}}$                                         \\
                                           & Idle          &  -                   & $p$                       & $\bar{n}\kappa_1T_{\text{meas}}$                                         \\ 
\midrule
\multirow{4}{*}{CNOT}                        & $Z_c$       &  -                   & $p$/3                     & $\bar{n}\kappa_1T_{\text{CX}} + \pi^2/(64\bar{n}\kappa_2T_{\text{CX}})$ \\
                                             & $Z_t$         &  -                   & $p$/3                     & $0.5\bar{n}\kappa_1T_{\text{CX}}$                                        \\
                                             & $Z_cZ_t$      &  -                   & $p$/3                     & $0.5\bar{n}\kappa_1T_{\text{CX}}$                                        \\
                                             & Idle          &  -                   & $p$                       & $\bar{n}\kappa_1T_{\text{CX}}$                                           \\
\midrule
\midrule
                                           &               & \multicolumn{2}{c}{$p_{Z_L}^{\rm tot} = A d(B p)^{C\lfloor\frac{d+1}{2}\rfloor}$} & $p_{Z_L}^{\rm tot} = A d(B \kappa_1/\kappa_2)^{C\lfloor\frac{d+1}{2}\rfloor}$                                        \\ 
\midrule
\multirow{3}{*}{\shortstack{$\text{Repetition}$ \\ $\text{code}$}}  &   $A$         &       0.32              &            0.12                &          0.07                 \\
                                                                    &   $B$    &         6.2            &              23              &         486                  \\
                                                                    &   $C$         &        1             &               0.99             &            0.94               \\
\midrule
\multirow{3}{*}{\shortstack{$\text{Cellular}$ \\ $\text{automaton}$ \\ $\text{codes}$ $\tetris$}}  &   $A$         &      0.07               &        0.019                    &              -             \\
                                                                    &   $B$    &             8.3        &           53                 &             -              \\
                                                                    &   $C$         &             0.99        &            0.95                &             -             \\
\midrule
\multirow{3}{*}{\shortstack{$\text{Optimized}$ \\ $\text{code}$ \\ $[136,34,22]^{\etoile}$}}  &   $A$         &     -               &       -                     &             0.1           \\
                                                                                    &   $B$         &             -        &           -                 &             1613             \\
                                                                                    &   $C$         &             -        &            -                &             0.94             \\
\bottomrule
\end{tabularx}
\caption{(Top) Physical error probabilities of the operations used in the stabilizer measurement circuits. The ``idle'' error probability corresponds to the phase-flip error probability applied to all the qubits that are idling while the corresponding operation is performed. (a) For a phenomenological error model, the phase-flip error applied to the data corresponds to an idling error with probability $p$ while the ancilla qubits are prepared in the $|+\rangle$ state. (b) For a generic phase-flip circuit-level noise, all of the operations have identical infidelity $p$. (c) The cat qubit circuit-level noise (see~\cite{Guillaud2019,Chamberland2022} for a detailed analysis) depends on the time of the operations $T_{\text{gate}}$, the single-photon loss rate $\kappa_1$ and the average photon number $\bar{n}$. In our simulation, we considered an average photon number $\bar{n} = 11$ and an identical time for all operations $T_{\text{prep}} = T_{\text{meas}} = T_{\text{CX}} = 1/\kappa_2$, where $\kappa_2$ is the two-photon dissipation rate. In this case, the errors are parametrized by the unique parameter $\kappa_1/\kappa_2$. (Bottom) Fitting parameters for the different codes and error models. The $\tetris$ cellular automaton codes with different distances are fitted together as they form a code family, while the optimized code is fitted independently (setting $A=0.1$). The fitting parameter $C$ is close to 1, indicating that the logical error probability is dominated by error configurations with exactly $\lfloor(d+1)/2\rfloor$ physical errors.
}
\label{tab:error_models}
\end{table*}
%\FloatBarrier
}

We now turn our attention to quantifying more precisely the logical error probability of some of the codes identified in the previous section, which exhibit the most favorable overhead reduction factor $kd/n$. Our focus here is on the performance of these codes as quantum memories, and we discuss the implementation of gates on these codes in the following section. The total logical error probability is given by $\epsilon_L \approx p_{X_L} + p_{Z_L}$ (as $p_{Y_L}=p_{X_L}p_{Z_L} \ll p_{X_L}, p_{Z_L} $), where $p_{X_L}$ and $p_{Z_L}$ are the probabilities of logical bit-flip and phase-flip errors per code cycle and per logical qubit, respectively. Since our phase-flip codes do not protect against bit-flip errors, we numerically estimate $p_{Z_L}$ only, and the logical bit-flip error is simply given by $p_{X_L} = N_{\text{cat qubits}} \times p_X / k$,  where $N_{\text{cat qubits}}$  is the total number of cat qubits in the architecture (including ancilla qubits, since bit-flip errors in ancilla qubits propagate to data qubits through the stabilizer measurement circuits) and $p_{X}$ is the physical bit-flip error probability per error correction cycle.
We numerically calculate the logical error probability for the cellular automaton code family characterized by the unique stabilizer shape $\tetris$, as well as for some of the optimized codes of Table~\ref{tab:codes}, and compare their performance to the repetition code family. We consider three error models, summarized in Table~\ref{tab:error_models}. The first model is a \textit{phenomenological} model (a) where each data qubit undergoes a phase-flip with probability $p$ before each error correction round, and each syndrome is read erroneously with an error probability $q=p$. This model allows for the comparison of intrinsic features of the codes, such as the threshold or the scaling of the logical error below threshold (\textit{i.e.}\ the scaling of the exponential suppression of errors with code distance). Next, we study the codes with a \textit{generic circuit-level} error model (b), where all code operations have a generic phase-flip error model parameterized by the operation infidelity $p$, assumed identical for all operations. This error model captures the effect of syndrome extraction circuits, such as the impact of the weight of the stabilizers. Finally, we simulate the codes using a \textit{cat qubit circuit-level} error model (c), which includes errors at the same locations as the generic circuit-level model, but where the precise error models are obtained using master equation simulations of noisy operations on cat qubits~\cite{Guillaud2019,Chamberland2022} parameterized by the relevant ratio $\kappa_1/\kappa_2$, where $\kappa_1$ is the single-photon loss rate of cat qubits and $\kappa_2$ is the two-photon stabilization rate (see Section~\ref{sec:Experiment} for more details). In this model, the different types of phase-flip errors do not happen with the same probability, in particular, the CNOT gate infidelity is dominated by ancilla phase-flip errors, which correspond to syndrome measurement errors. This error model provides the most realistic estimate of the performance of the architecture when concatenated with cat qubits.

Regardless of the error model, we estimate the logical error as follows. For a code of distance $d$, we simulate $d$ rounds of stabilizer measurements, followed by a perfect stabilizer measurement to ensure convergence in the codespace. The $d$ rounds are decoded using the belief propagation + ordered statistics decoding (BP+OSD) decoder~\cite{panteleev2021,roffe2020}, adapted to the case of multiple temporal rounds following~\cite{bravyi2023}, using the library~\cite{Roffe_LDPC_Python_tools_2022} (more details about the decoding parameters are given in Appendix~\ref{sec:bp+osd}). The decoder succeeds if the proposed correction successfully removes the errors, otherwise it introduces a logical error. The circuit is sampled until $100$ logical errors are observed, and the probability of a logical phase-flip error per logical qubit after $d$ rounds is estimated as $p_{Z_L}^{\rm tot} = 1 - (1 - 100/N)^{1/k}$ where $N$ is the total number of samples. The logical phase-flip error per round and per logical qubit is then estimated as $p_{Z_L}^{\rm tot} \approx dp_{Z_L}$. In the asymptotic regime ($p$ or $\kappa_1/\kappa_2$ $\ll 1$), the logical error probability is dominated by error configurations where there are exactly $\lfloor (d+1)/2 \rfloor$ errors. Therefore, the total error probability $p_{Z_L}^{\rm tot}$ after $d$ rounds is fitted to the ansatz~\cite{Chamberland2022}

\begin{equation}\label{eq:logical_ansatz}
    p_{Z_L}^{\rm tot} = A d(B p)^{C\lfloor\frac{d+1}{2}\rfloor}
\end{equation}
or
\begin{equation}\label{eq:logical_ansatz_cat}
    p_{Z_L}^{\rm tot} = A d(B \kappa_1/\kappa_2)^{C\lfloor\frac{d+1}{2}\rfloor}. 
\end{equation}

The fitted parameters $(A, B, C)$ are summarized in Table~\ref{tab:error_models}. We plot the logical error probability as a function of physical noise error probability in Figure~\ref{fig:numerics} for (a) the phenomenological model with $p = q$, (b) the generic circuit-level model, and (c) the circuit-level model of cat qubits. Initially, we observe in Figure~\ref{fig:numerics} (a) that for a phenomenological model, the codes appear to have the same threshold and the same scaling below the threshold, since the curves with the same distance are almost perfectly superimposed. Note in particular that the cellular automaton code with different stabilizer shapes in each row has the same logical error as the repetition or single-shape cellular automaton code of identical distance. This indicates that the appropriate metric characterizing code performance is the distance, and that the weight or shape of the stabilizers does not play a significant role in the case of cellular automaton codes. In Figure~\ref{fig:numerics}(b), we see that, under generic circuit-level phase-flip noise, the threshold of cellular automaton codes is divided by a factor two. We attribute this to the fact that the syndrome measurement circuit is roughly twice deeper for these codes, as the weight-four stabilizers require four CNOT gates to be measured instead of two for the weight-two repetition code. Finally, in Figure~\ref{fig:numerics}(c), we represent the logical phase-flip as a function of $\kappa_1/\kappa_2$ at a fixed number of photons. The number of photons is chosen such that the bit-flip is sufficiently small to achieve an overall logical error around $10^{-8}$ per error correction cycle. Indeed, by extrapolating the fit corresponding to $\bar{n} = 11$ photons of Table~\ref{tab:error_models} to $\kappa_1/\kappa_2 = 10^{-4}$, we estimate that a logical phase-flip error probability per cycle and per logical qubit $p_{Z_L} = 6.4\times 10^{-10}$ can be achieved with the $[136,34,22]$ code. This logical error phase-flip error probability is roughly identical for the $[165+8\ell, 34+2\ell, 22]$ code family (the extended version of the $[136,34,22]$ code without periodic boundary conditions as described in Section~\ref{sec:Experiment}), which we use here to encode $100$ logical qubits in the $[429,100,22]$ code. The logical bit-flip probability is dominated by the errors occurring during $\text{CNOT}$ gates $p^{\text{CX}}_X = 0.5e^{-2\bar{n}}$~\cite{LeRegent2023highperformance,Gouzien2023}, thus the probability of logical bit-flip per cycle at 11 photons is given by $p_{X_L} = N_{\text{CX}} \times p^{\text{CX}}_X / k = 1.8\times 10^{-9}$, where $N_{\text{CX}}$ is the number of CNOT gates performed during an error correction cycle. Thus, the code is expected to achieve a total error probability $\epsilon_L = 2.5\times10^{-9}$, with $429$ data cat qubits, for a total number of $758$ cat qubits. Note that a lower value of $\kappa_1/\kappa_2$ would enable the use of even lower code distances, thereby further increasing the encoding rate. However, the number of physical data cat qubits per logical qubit of this code is $4.3$ (for a code of distance $22$), such that this code is already extremely efficient.

\section{\label{sec:gates}Fault-tolerant implementation of a universal set of logical gates}

%-------------------------------%
\begin{figure}[t!]
    \centering
    \includegraphics[width=\columnwidth]{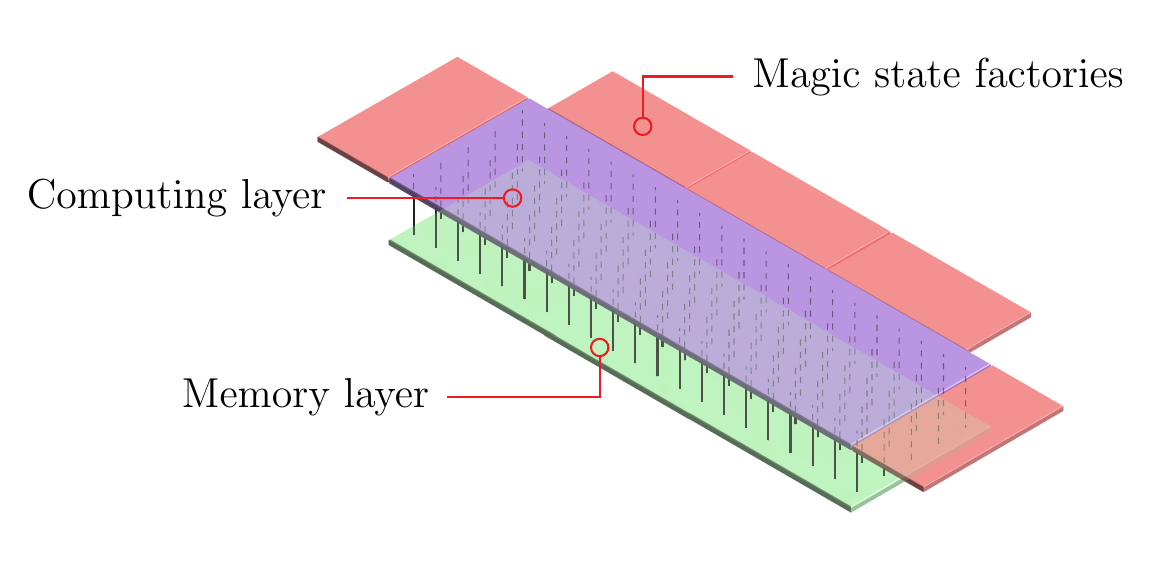}
    \vspace{-0.5cm}
    \caption{\label{fig:3D_processor}
    Example of a layout for a quantum computer. The ancillary logical qubits of the upper ``compute'' layer are encoded in repetition codes, while the algorithmic logical qubits of the lower ``memory'' layer are encoded in high-rate phase-flip codes.
    }
\end{figure}
%-------------------------------%

``Block'' codes, where the coding blocks contain several logical qubits, allow for a compact encoding, but the price to pay is that it is more difficult to manipulate logical information.
Indeed, unlike the case where only one logical qubit is encoded per block, here the supports of the logical operators corresponding to different logical qubits overlap, such that each physical qubit belong to the support of several logical operators of distinct logical qubits.
Therefore, it is not trivial to individually address logical qubits: while Pauli operations can be directly applied to the code by performing the corresponding qubit gate on the support of the operator, it is not always clear how to measure these operators or to perform multi-qubit gates within a single block.

%-------------------------------%
\begin{figure*}[t!]
    \centering
    \includegraphics[width=\textwidth]{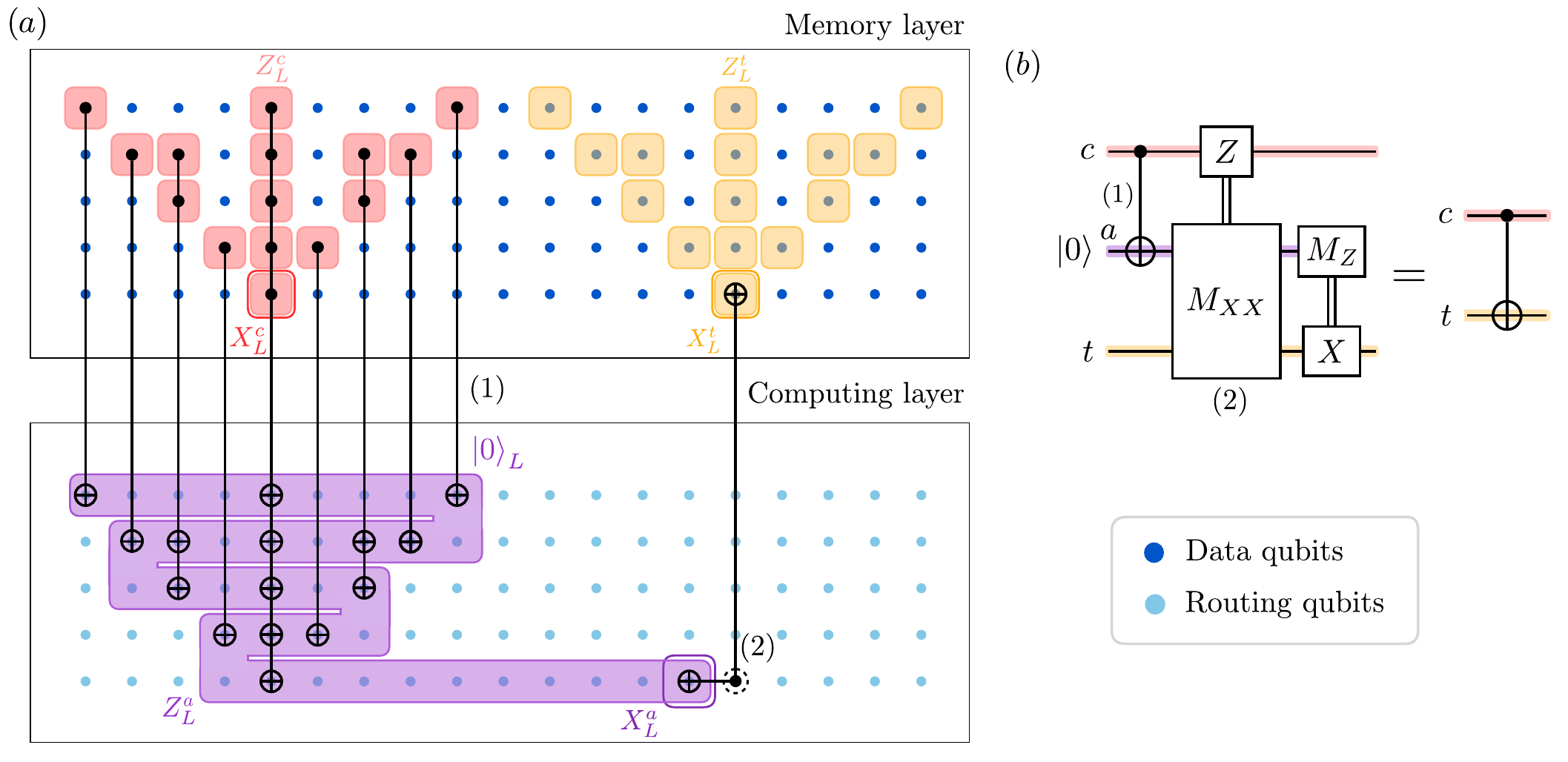}
    \vspace{-0.5cm}
    \caption{\label{fig:cnot}
    Logical CNOT gate between two logical qubits encoded in the cellular automaton code. Figure (b) presents the lattice surgery circuit to be implemented~\cite{Horsman2012}. As depicted in Figure (a), the implementation of the lattice surgery scheme requires routing qubits encoded in the repetition code. This routing circuitry can be seen as a second layer on top of the LDPC block, where CNOT are allowed between corresponding qubits in the LDPC and repetition code block. First, a logical $\ket{0}_L$ is prepared. A logical CNOT gate is performed between the control logical qubit and the logical $\ket{0}_L$. This is followed by an $M_{XX}$ measurement between the logical $\ket{0}_L$ and the target logical qubit. This process includes preparing a routing ancilla (dashed circle) in the $\ket{+}$ state, performing two CNOTs with the logical $\ket{0}_L$ and the logical $X_L^t$ of the target logical qubit, and finally measuring the ancilla. The result is the $M_{XX}$ measurement and the procedure is repeated $d$ times to ensure fault-tolerance. Finally, the logical ancilla is measured in the $Z$ basis (by measuring individual qubits in $Z$ and taking the product) and some Pauli corrections are applied depending on the $M_{XX}$ and $M_{Z}$ measurements.
    }
\end{figure*}
%-------------------------------%

The initial proposals for quantum computing with information encoded on multi-qubit codes were based on teleportation: the logical qubit to be manipulated is extracted from the code block by teleportation to another block where it is isolated to apply some logical gate, before being reinserted into its original place by teleportation~\cite{gottesman1998b,steane1999}. 
In order to reduce the overhead of these protocols, more recent works propose to use state preparation and injection~\cite{steane2005,gottesman2013} or, more efficiently, to adapt the lattice surgery techniques~\cite{cohen2022,cowtan2023} developed for the surface code~\cite{Horsman2012}, where operations are carried out through stabilizer measurements, or to perform code deformation methods~\cite{breuckmann2017,krishna2021}.

Here, we propose to implement a universal set of logical gates on our classical LDPC codes with a two-layer architecture, which can be realized with existing technologies, as discussed in more detail in Section~\ref{sec:Experiment}.
The lower layer is a ``memory'' layer that contains the logical qubits encoded in high-rate local phase-flip codes, and the upper layer is a ``computing'' layer that contains logical routing qubits and magic state factories encoded in repetition codes, as depicted in Figure~\ref{fig:3D_processor}.
Each of the data qubits of the memory layer is connected to the corresponding data qubit of the computing layer, in order to implement physical CNOT gates.
Note that this construction only increases the connectivity of the logical data qubit of the memory layer by one while staying local.
The ancillary logical qubits encoded in repetition codes are used to implement a universal set of logical gates on the qubits of the memory layer by adapting the schemes proposed for the repetition code architecture~\cite{Guillaud2019,Chamberland2022,Gouzien2023}.
More precisely, the set of logical states $\{|+i\rangle_L, |\text{C}Z\rangle_L, |\text{CC}X\rangle_L\}$ can be prepared using fault-tolerant logical measurements on repetition codes~\cite{Chamberland2022}. This set of states allows for a universal gate set when supplemented with the set of operations $\{\mathcal{P}_{|0\rangle_L}, \mathcal{P}_{|+\rangle_L}, Z_L, X_L, \text{C}X_L, \mathcal{M}_{Z_L}, \mathcal{M}_{X_L}\}$, some of which can be directly realized in the memory layer, as we now detail. In the remainder of this section, we focus on cellular automaton codes for which the support of logical Pauli operators is easily determined but the schemes are general and can be realized for all of the codes of Figure~\ref{fig:AllCodes}. As discussed in Section~\ref{sec:codessearch}, the support of logical $Z_L$ operators can be obtained by applying the cellular automaton rules corresponding to the stabilizers. In this case, the logical $X_L$ operators have weight one, and correspond to the physical $X$ operators of the lowest $m-1$ rows.
 
\textit{Pauli measurements.} 
% Destructive measurements
The logical qubits can be simultaneously destructively measured in the logical $Z_L$ basis (typically, at the end of the algorithm) by measuring all of the data cat qubits in the $Z$ basis. The value of each of the logical $Z_L$ operators is then obtained by multiplying all the measurement outcomes corresponding to the qubits in their support. Here, the fault-tolerance is guaranteed from the fact that only an (exponentially suppressed) bit-flip error can flip the logical measurement outcome.
Similarly, the logical qubits can be simultaneously destructively measured in the logical $X_L$ basis by measuring all of the data cat qubits in the $X$ basis. The results of the measurement outcomes are used to infer the value of the stabilizers, which are fed to the decoder. The final measurement outcomes of the $X_L$ operators are then deduced from the measurement outcomes of the $(m-1)$ bottom rows by flipping the measurement outcomes of the qubits where the decoder predicted a phase-flip error.
Note that it is not possible to simultaneously destructively measure all of the logical qubits in different bases in a fault-tolerant manner, as the physical qubits of each logical qubit measured in the logical $Z_L$ basis are measured in the $Z$ basis, such that the corresponding measurement outcomes may no longer be used to infer the value of the stabilizers in which they participate.

% QND measurements
Alternatively, the logical Pauli operator $Z_L$ of any logical qubit in the memory layer can be measured in a quantum non-demolition (QND) manner using the computing layer by performing a logical transversal CNOT between the LDPC logical qubit as control and an ancillary logical qubit as target of the computing layer prepared in $|0\rangle_L$, followed by a logical $Z_L$ measurement of this ancillary qubit. The logical Pauli operator $X_L$ is measured by performing a physical CNOT gate between the support of the weight-one $X_L$ operator of any logical qubit in the memory layer as target and an ancillary physical qubit of the computing layer, prepared in a $\ket{+}$ state, as control. The ancilla is then measured in the $X$ basis, and this procedure is repeated $d$ times to ensure fault-tolerance. Finally, the $X_L$ measurement value is deduced with a majority vote.

\textit{Pauli preparations.} 
%Reformulation mineure proposée, à toi de voir ce que tu préfères
While the logical $|0\rangle_L$, $|+\rangle_L$ states could be prepared using the QND measurement of $Z_L$ and $X_L$, the standard surface code method~\cite{Fowler2012} that leverages stabilizer measurements can be used without resorting to the computing layer. To prepare all of the logical qubits in an eigenstate of the logical $X_L$ operator, each physical qubit is prepared in either $\ket{+}$ or $\ket{-}$ accordingly (a state where all logical qubits are in an eigenstate of the logical $X_L$ operator is separable). Similarly, to prepare all of the logical qubits in an eigenstate of the logical $Z_L$ operator, each of the corresponding physical qubit in the lowest $m-1$ are prepared in $|0\rangle$ or $|1\rangle$, depending on the target eigenstate, and all of the qubits of the other rows are prepared in the $|0\rangle$ state. Note that this (separable) state is an eigenstate of all the logical $Z_L$ operators with the desired eigenvalues, but is not in the code space. Therefore, $d$ rounds of stabilizer measurements are then performed (followed by a correction that may be tracked in software) to project the state in the code space, which concludes the state preparation.
Note that, symmetrically to measurements, it is not possible to prepare eigenstates of different Pauli operators simultaneously.

\textit{Intra-block $\text{C}X_L$ gate.}
Figure~\ref{fig:cnot}(a) shows how the computing layer may be used to directly implement a logical CNOT gate between two logical qubits of the memory layer with the lattice surgery scheme of~\cite{Horsman2012}, depicted on Figure~\ref{fig:cnot}(b). First, a logical $|0\rangle^a_L$ ancilla state is prepared in a repetition code of the computing layer on qubits that cover all of the support of the logical $Z^c_L$ operator of the control, up until the neighbouring qubit of the corresponding $X^t_L$ operator of the target. Then, a logical CNOT gate is performed transversally between the two layers, followed by a QND logical $X^a_LX^t_L$ and a logical $Z^a_L$ measurement. Pauli corrections conditioned on the logical measurement outcomes complete the gate teleportation. Note that this scheme can be extended to realize multiple-target controlled NOT gate $\text{C}X_L^k$~\cite{Gouzien2023}, or several CNOT gates sharing the same target in parallel, but that it is not possible to apply simultaneously two logical CNOT gates with different target logical qubits if their respective logical control qubit overlap.
As a consequence, some level of parallelization is possible, but with some constraints.
A full assessment of the slowdown incurred when compiling circuits on this architecture remains to be done.

\textit{Magic state injection.}
Finally, to complete the universal logical gate set, we show how to inject logical magic states in the memory layer.
The states are prepared in dedicated magic state factories in the computing layer as depicted in Figure~\ref{fig:3D_processor}.
The injection requires either to perform CNOT gates with the magic state as a control and the logical qubit encoded in the memory layer as a target or vice versa. This can be achieved with minor adaptations of the techniques used for the intra-block $\text{C}X_L$ gate (Figure~\ref{fig:Toffoli}) and the detailed operations are presented in Appendix~\ref{sec:state_injection}.

%-------------------------------%
\begin{figure}[t!]
    \centering
    \includegraphics[width=\columnwidth]{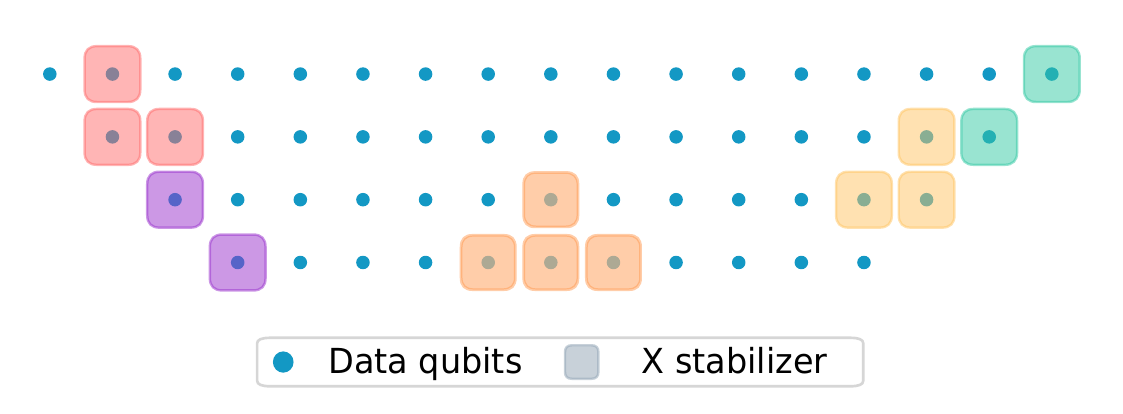}
    \vspace{-0.5cm}
    \caption{\label{fig:bord}
    Cellular automaton code without periodic boundary conditions. The number of logical qubits encoded still corresponds to the number of physical qubits in the bottom row, but the lattice is extended on both sides to enable the logical qubits on the side to have the same distance. The stabilizer which overlaps with the edges are simply truncated.
    }
\end{figure}
%-------------------------------%

%-------------------------------%
\begin{figure*}[t!]
    \centering
    \includegraphics[width=\textwidth]{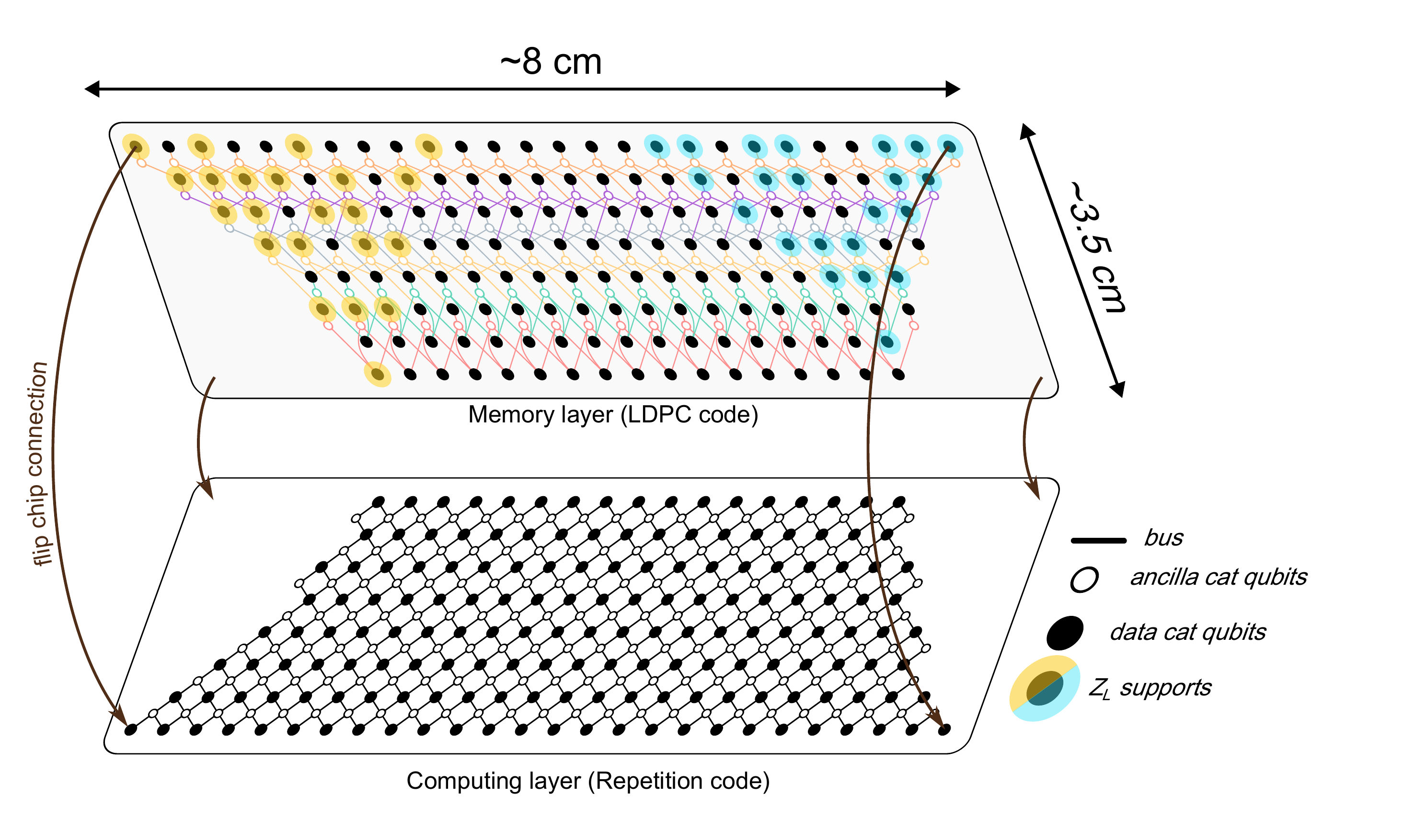}
    \vspace{-0.5cm}
    \caption{\label{fig:chiplayout}
    Layout of a $[165,34,22]^{\etoile}$ cellular automaton code corresponding to the optimized code of Table~\ref{tab:codes} for $H = 8$. The architecture comprises a memory layer where the information is encoded in the LDPC code and a repetition code computing layer enabling logical gates between qubits encoded in the memory layer.
    Flip chip technology enables connections between corresponding qubits in the memory layer and the computing layer.
    The weight-4 stabilizers of the code features crossing cables but this can be circumvented by passing through the other plane.
    The trapezoidal shape ensures that the support of the $Z_L$ logical operators on the side (in blue and yellow) achieve full distance in order to preserve the code parameters.
    All data qubits and ancilla qubits are represented (except for the magic state factories) for a total of 296 qubits for the memory layer and 311 for the computing layer.}
\end{figure*}
%-------------------------------%

\section{\label{sec:Experiment}Towards experimental implementation}

In this section, we provide practical details on the experimental implementation of our architecture. We begin by explaining how to remove periodic boundary conditions, before discussing how the chip can be realized with already available technologies. Finally, we discuss recent experiments with dissipative cat qubits in order to provide context to the hardware assumptions used in this work.

\textit{Removing periodic boundary conditions.} 
In Section~\ref{sec:codessearch}, the codes were searched by imposing periodic boundary conditions on the lateral edges, in order to avoid having to consider the precise shape of the code boundaries, which depend on the shape of the stabilizers. For a practical 2D implementation, however, it is preferable to remove these periodic conditions to preserve the locality of the code.
Nevertheless, it is essential to ensure the codes are made planar without compromising the parameters $[n,k,d]$. This can be done in the case of cellular automaton code as illustrated in Figure~\ref{fig:bord} by adding data qubits on both edges, in order to preserve the support of the logical operators $Z_L$ which previously ``looped'' to the opposite side of the chip. The stabilizers at the edge are then simply truncated when they exceed the grid of physical qubits. While this requires adding a few physical qubits in order to maintain the code distance, this overhead is negligible in practice for the typical regime of interest ($d \approx 10-30$ and $k \approx 100-10,000$). Indeed, for a cellular automaton code, the number of qubits added on the sides is on the order of $H^2$, independent of the lattice width $L$ which allows increasing the number of logical qubits.

\textit{Chip layout.}
Thanks to the fact that we are considering local codes in 2D, the physical realization of the chip is greatly simplified compared to architectures that use high-encoding rate quantum LDPC codes for standard qubits, which necessarily require long-range connectivity.
Here, the codes can be implemented and operated within a single memory layer, with local stabilizers of weight 4, that is, with exactly the same constraints as the surface code that has already been experimentally realized~\cite{google2023}.
Note that the locality of the codes alleviates the need for long-range couplers, which are technologically challenging to realize in large-scale architectures.
To operate the architecture, as discussed in Section~\ref{sec:gates}, it suffices to add a computing layer with repetition codes, which remains simpler than many existing proposals to realize gates on high-rate LDPC codes~\cite{Tremblay2022,bravyi2023}.
A possible way to realize our two-layer architecture is to use flip-chip technology, inspired by semiconductors~\cite{Pltner1991} and successfully adapted to superconducting processors~\cite{Rosenberg2017,Foxen2017}.
The two layers are manufactured separately and then joined face-to-face using indium bump-bonds, where the indium establishes a superconducting galvanic connection between the two chips.
Alternatively, the technology of TSV (through silicon vias)~\cite{Yost2020} would also allow for the realization of our architecture (both sides of a chip are metallized and connectivity is established through the substrate).
Figure~\ref{fig:chiplayout} represents all qubits, including ancillary qubits and routing qubits, and the required connectivity to implement and operate the $[165, 34, 22]$ code (the qubits of the magic state factories of the computing layer are not shown).
Note that, although some lines cross, this is not a problem here. Indeed, as the architecture is realized on two planes, a line can pass from one plane to another to avoid crossings.
What matters then is the density of crossings per unit cell (for footprint-on-chip reasons), which is here constant thanks to the locality of the codes.

\textit{Current state of the experimental art and further research directions.} 
The theoretical estimation of the performance of our codes relies on two independent hardware assumptions. The first one is about the minimal physical bit-flip error probability that can be achieved thanks to the exponential suppression of bit-flip errors with the average photon number $\bar n=|\alpha|^2$. Assuming for instance that $\kappa_2/2\pi = \SI{3.18}{\mega\hertz}$, the average photon number $\bar n = 11$ corresponds to a cat qubit bit-flip time of $T_X = 13$ minutes.
The second one is about the value for the ratio $\kappa_1/\kappa_2$, where $1/\kappa_1$ is the single-photon lifetime of the resonator hosting the cat qubit and $\kappa_2$ the two-photon stabilisation rate of the cat qubit. This ratio determines the phase-flip error models of all cat qubit operations.  As a reminder, for $\kappa_1/\kappa_2=10^{-4}$, the $[429,100,22]$ code can be used to store one hundred logical qubits with logical error probability $\epsilon_L \leq 10^{-8}$, with an average photon number in the cat states $\bar n = 11$ (the number 758 of physical cat qubits given in Table~\ref{tab:architectures_comparison} includes the measurement ancilla qubits).  We now review recent cat qubit experiments to put these numbers in perspective.

\textit{Experimental progress towards macroscopic bit-flip times.} Loosely speaking, as long as the error channels of the harmonic oscillator encoding the cat qubit induce local errors in phase space, the increasing separation of the states $|\alpha\rangle$ and $|-\alpha\rangle$ with the average number of photons allows for the exponential suppression of the resulting bit-flips at the encoding level. The bit-flip rate saturates at the rate of the first physical phenomenon that induces non-local errors in the phase space. This is somewhat similar to the discrete-variable error correcting codes, like the surface code: the exponential suppression of logical errors with the code distance assumes that errors are acting locally and independently on each physical qubits. In experiments, this exponential suppression of logical errors is indeed observed until it saturates at a rate corresponding to the dominant physical error channel that induces spatially correlated errors over many qubits (in the case of~\cite{google2023}, this is due to high-energy cosmic rays). The first experimental demonstrations of the stabilization of cat states with two-photon dissipation~\cite{Leghtas2015,Touzard2018} did not show any exponential bit-flip suppression. In these experiments, the two-photon exchange medium was a parametrically driven transmon-type circuit. These experiments suffered from strong parasitic interactions (stronger than the engineered two-photon dissipation) that led to non-local perturbations, and prevented  any bit-flip suppression. The subsequent experiments replaced this two-photon exchange circuit with an Asymmetrically Threaded SQUID (nicknamed ATS) that removed most of these parasitic interactions, and demonstrated the exponential suppression of bit-flips by up to a factor of 300~\cite{Lescanne2020}. In this experiment, the suppression of bit-flips saturated around 5 photons, achieving a bit-flip lifetime of $T_X = \SI{e-3}{\second}$ (for a phase-flip rate at 5 photons of $T_Z = \SI{5e-7}{\second}$). The saturation of the bit-flip time was attributed to strong dispersive coupling with a thermally excited transmon (used for the quantum state tomography of the cat state). A follow-up experiment confirmed that this saturation was indeed due to the transmon~\cite{Berdou2022} and that by removing it, macroscopic bit-flip times of  $100$ seconds should be achievable.  This was demonstrated in a recent experiment~\cite{reglade2023}, where a bit-flip time of $T_X=\SI{15}{\second}$ was observed for a cat qubit with $\bar{n}=11.3$ photons (for a corresponding phase-flip rate of $T_Z=\SI{4.9e-7}{\second}$).

The exact physical error channel causing the bit-flip saturation at a scale of a few tens of seconds remains unclear. However, the impact of high-energy cosmic rays causing non-local errors on the superconducting chip was identified in surface code experiments as the limiting factor for the exponential suppression of logical errors~\cite{mcewen2022resolving}. The impact of such events on cat qubits is not yet clear, but the time scales involved are similar, which suggests that it might be the limiting factor.

\textit{Experimental progress for $\kappa_1/\kappa_2$.} In order to decrease $\kappa_1/\kappa_2$, two complementary approaches may be used: increase $\kappa_2$ and decrease $\kappa_1$.

We first summarize recent progress to increase $\kappa_2$. Using 2D microwave resonators, the initial experiment that demonstrated the stabilization of a cat state achieved $\kappa_1/\kappa_2$ ratios close to 1 ($\kappa_1/2\pi = \SI{53}{\kilo\hertz}$, and $\kappa_2/2\pi = \SI{40}{\kilo\hertz}$ in Lescanne \textit{et al.}~\cite{Lescanne2020}). Since this work, progress in cat qubit circuit engineering has allowed to increase this ratio by more than two orders of magnitude to achieve more recently a ratio $\kappa_1/\kappa_2 = 6.5\times 10^{-3}$ ($\kappa_2/2\pi = \SI{2.16}{\mega\hertz}$, $\kappa_1/2\pi = \SI{14}{\kilo\hertz}$)~\cite{marquet2023}. These circuits may be further optimized, and there is no fundamental theoretical reason to think that one cannot achieve higher values for $\kappa_2$, perhaps in the range $\kappa_2/2\pi \geq \SI{10}{\mega\hertz}$. Therefore, we deem a value around $\kappa_1/\kappa_2 = 10^{-4}$ may be achieved when the platform reaches its experimental maturity.

In order to achieve lower values for $\kappa_1$, besides progress in material science or general microwave packaging, a different approach involves using a different resonator technology. This could be achieved \textit{e.g.}\ by using 3D resonators whose lifetimes can reach up to $\SI{34}{\milli\second}$ (\cite{Milul2023}, corresponding to a $\kappa_1/2\pi = \SI{4.7}{\hertz}$). Another approach would be to use acoustic resonators coupled to superconducting circuits (see~\cite{Chamberland2022} for an analysis with cat states where $\kappa_1/\kappa_2 = 10^{-5}$ is proposed with $\kappa_1/2\pi = \SI{2.8}{\hertz}$, $\kappa_2/2\pi = \SI{280}{\kilo\hertz}$). For such a change of resonator technology, the layout of Fig.~\ref{fig:chiplayout} with 2D superconducting resonators would need to be adapted accordingly. 

While these recent experimental works on dissipative cat qubits have characterized the bit-flip time for an idling cat qubit, it was demonstrated in~\cite{reglade2023} that the macroscopic bit-flip times were preserved while rotations around the $Z$ axis of the Bloch sphere were performed on the cat qubit. An important assumption that our architecture relies on is that all of the physical gates applied to the cat qubit preserve the noise bias, which has yet to be experimentally demonstrated, \textit{e.g.}\ for the two-qubit CNOT gate, although recent progress towards this goal has been made~\cite{cottet2023towards}. Furthermore, the demonstration of robust suppression of bit-flips in a multi-cat qubit chip also needs to be done.

\section{\label{sec:conclusion}Conclusion}
We proposed and analyzed an architecture for a fault-tolerant quantum computer based on the concatenation of cat qubits in 2D-local phase-flip LDPC codes with high encoding rates.
We found that the combination of cat qubits, which simplify the task of quantum error correction due to their robustness against bit-flip, with classical LDPC codes allowed for the realisation of logical qubits with an extremely low hardware overhead.
More precisely, assuming a ratio between the rates of single-photon loss and two-photon stabilisation $\kappa_1/\kappa_2 = 10^{-4}$, we showed how a processor with $100$ logical qubits used as a memory with a logical error probability per cycle and per logical qubit of $\epsilon_L \leq 10^{-8}$ could be built using only $758$ cat qubits, a reduction by a factor of $2.8$ compared to the architecture based on repetition codes.

Remarkably, we have been able to construct a completely local architecture in 2D, with low-weight stabilizers.
Indeed, while it is well known that quantum LDPC codes with better encoding rates than the surface code require non-local connectivity in 2D, this result does not apply to classical codes local in 2D, which enabled us to construct 2D local codes with much higher encoding rates than the repetition code.
Moreover, the realization of a universal set of logical gates on these phase-flip codes is also simpler than previous schemes that implement gates on high-rate qLDPC codes.
We have shown how a universal set of logical gates could be achieved with a computing layer containing ancillary logical qubits encoded in repetition codes, without sacrificing locality and with only a slight increase in connectivity.
This shows that, while cat qubits have been appreciated so far for the simplification of the active error correction layer they allow, they also drastically simplify other aspects of a fault-tolerant architecture.
Thus, the ``technological cost'' of the proposed architecture is similar to the surface code and could already be realized with currently available technology, but it allows for reducing by a factor 44 (compared to the surface code) the number of physical qubits necessary to implement $100$ logical qubits with $10^{-8}$ logical error probability, used as quantum memories.

One may wonder whether this reduction in overhead for storing the logical quantum information is maintained when computing, since the logical qubits of the computing layer are encoded in repetition codes (note that these qubits only roughly double the total number of cat qubits). In order to quantify more precisely how the gains in ``memory density'' reduce the amount of resources needed to run a fault-tolerant quantum algorithm, we benchmark our architecture against the repetition cat code architecture for which a recent theoretical work~\cite{Gouzien2023} estimated that the factorization of RSA-2048 integers could be theoretically done in 4 days using $\num{350000}$ cat qubits. We estimate that, under the same hardware assumptions ($\kappa_1/\kappa_2=10^{-5}$), the improvements proposed in this paper would reduce this number to less than $\num{100000}$ cat qubits, and 7 days of computation (see Appendix~\ref{sec:RSA_resource_estimation}).

These results indicate that the combination of cat qubits with classical LDPC codes produces a viable architecture for realizing a large-scale quantum computer at the cost of a manageable hardware overhead.
This architecture could be further improved by optimizing protocols for implementing logical gates or by continuing to improve the design of cat qubits at the hardware level.
For instance, we estimate that the assumption of $\kappa_1/\kappa_2 = 10^{-4}$ could be relaxed due to the rapid advancement of theoretical research on cat qubits and the improvement of the architecture at the lower hardware levels, as discussed in Appendix~\ref{subsec:archi_cat}.
Alternatively, this could allow for a reduction in the distance of the codes, thereby further improving the encoding rate, or reduce the logical error at constant overhead to approach the error rates required for the most challenging applications.
A different approach could be to replace the two-component cat qubits by higher-dimensional cat states, as the type of codes we use are known to be more efficient with qudits~\cite{yoshida2013}, although such cat qudits are experimentally more challenging to stabilize.
Finally, the construction of an efficient decoder for our codes that could be integrated into the architecture of the quantum processor to decode syndromes in real-time is still an open research problem.

\begin{acknowledgments}
D.R and J.G would like to thank Paul Magnard for countless discussions about the feasibility of our proposal, for helping with the Figures, and for his enthusiasm for the final architecture. All authors thank Barbara Terhal, Benjamin Brown and Theodore Yoder for their comments on the manuscript, and \'Elie Gouzien for computing the resources needed to factor a 2048-RSA integer on our architecture and for suggesting and helping with the use of SAT solvers.
We acknowledge funding from the Plan France 2030 through the project ANR-22-PETQ-0006. Mazyar Mirrahimi thanks funding from QuantERA grant QuCOS, by ANR 19-QUAN-0006-04. 
\end{acknowledgments}

%%%%%%%%%%%%%%%%%%%%%%%%%%%%%%%%
% APPENDICES
%%%%%%%%%%%%%%%%%%%%%%%%%%%%%%%%

\appendix

\section{\label{sec:archi_comp}Architectures comparison}

Comparing architectures that do not rely on the same types of qubit is a somewhat difficult task because one cannot directly use the same error models to make an apples-to-apples comparison. However, it remains an interesting exercise that sheds light on the importance of the assumptions made for each architecture, and allows for a better understanding of the significance of experimental progress for each physical platform. We first detail how the figures in Table~\ref{tab:architectures_comparison} were obtained, before discussing several recent theoretical proposals to further optimise the performance of the cat qubit architecture.

\subsection{\label{subsec:archi_footprints}Footprints}

\textit{Surface code.} Assuming a circuit-level depolarizing noise where all of the operations in the syndrome measurements circuits are noisy, with identical infidelity $\epsilon$, and using a minimum-weight perfect matching decoder, the logical error probability per code cycle $\epsilon_L$ is well approximated by~\cite{Fowler2013,fowler2019low}
\begin{equation}
    \epsilon_L = 0.1(100\epsilon)^{\lfloor \frac{d+1}{2} \rfloor}.
\end{equation}
For $\epsilon=10^{-3}$, a logical error rate $\epsilon_L = 10^{-8}$ is achieved for a code distance $d=13$. 
With these numbers, one can encode $N_L=100$ logical qubits with $N = N_L (2d^2-1) = 33,700$ physical qubits (including ancilla qubits).

\textit{Small qLDPC code.} We focus for this example on the $[[144, 12, 12]]$ qLDPC code introduced in~\cite{bravyi2023}. This code is a quasi-cyclic code of CSS type with stabilizer operators of weight 6.
The Tanner graph of this code has thickness two, which implies from a technological point of view that the code can be implementable using two planer layers of couplers that do not intersect (see~\cite{bravyi2023} for a detailed discussion).
Under a circuit-level depolarizing noise of strength $\epsilon$ and using a BP+OSD decoder, the logical error probability per code cycle and per logical qubit $\epsilon_L$ is well approximated by~\cite{bravyi2023}
\begin{equation}
    12\epsilon_L = \epsilon^5 e^{c_0 + c_1\epsilon + c_2\epsilon^2},
\end{equation}
where $c_0 = 16.46, c_1 = 1076, c_2 = -54522$. Thus, $\epsilon=10^{-3}$ corresponds to a logical error rate $\epsilon_L = 3.3\times10^{-9}$ and the total number of physical qubits required to implement $N_L = 100$ logical qubits is $N = 2nN_L/12 = 2,400$. 

\textit{Cat qubits and repetition codes.}
Assuming all of the operations of the repetition code (ancilla cat state preparation, CNOT gates, and ancilla measurement, see~\cite{Mirrahimi2014,Guillaud2019} for a detailed review) are realized in time $T = 1/\kappa_2$, where $\kappa_2$ is the two-photon dissipation rate, and assuming that the only error channel is single-photon loss at a rate $\kappa_1$, the logical error probability per code cycle $\epsilon_L$ is well approximated by~\cite{Gouzien2023}
\begin{equation}\label{eq:logical_error_repcat}
    \epsilon_L = 5.6\times10^{-2}{\left( \dfrac{{(\alpha^{2})}^{0.86}\kappa_1/\kappa_2}{1.3\times10^{-2}} \right)}^{\frac{d+1}{2}} + 2(d-1)\times 0.50 e^{-2\bar{n}},
\end{equation}
where $d$ is the repetition code distance (against phase-flip) and $\bar{n}$ is the mean photon number of the cat, playing the role of the ``distance'' against bit-flips. Note that here $\epsilon_L = \epsilon^Z_L + \epsilon^X_L$ corresponds to the total logical error rate, while in the two previous examples $\epsilon_L$ is the logical error probability of either logical bit-flips or phase-flips (identical).
Assuming a ratio $\kappa_1/\kappa_2 = 10^{-4}$, one achieves a logical error rate $\epsilon_L = 2.8\times10^{-9} + 2.7\times10^{-9} = 5.5\times10^{-9}$ using code distances $(\bar{n}, d) = (11, 11)$, which corresponds to a total number of cat qubits $N = (2d-1)N_L = 2,100$.

Note that these assumptions translate (see Table~\ref{tab:error_models}) into state preparation and measurement infidelities of $\epsilon_{\text{SPAM}} = 1.1\times10^{-3}$, CNOT gate infidelity of $\epsilon_{\text{CNOT}} = 1.6\times10^{-2}$ and idling errors of $\epsilon_{\text{idling}} = 1.1\times10^{-3}$, that is, all operations are noisier than for a depolarizing error model with strength $\epsilon = 10^{-3}$.

\subsection{\label{subsec:archi_cat}Theoretical optimisations}

The cat qubit architecture has also been the subject of many very recent theoretical proposals that have not yet been realized experimentally, but which are expected to potentially relax the constraint on $\kappa_1/\kappa_2$ by an order of magnitude. That is, to achieve figures close to those in Table~\ref{tab:architectures_comparison} for $\kappa_1/\kappa_2 = 10^{-3}$, or alternatively, to reach a logical error rate $\epsilon_L=10^{-13}$ with $\kappa_1/\kappa_2 = 10^{-4}$.

A first optimization is the use of asymmetric repetition codes~\cite{LeRegent2023highperformance}, whose performance is increased thanks to the ``specialization'' of cat qubits. More specifically, it has been demonstrated that the more relevant ratio is $\kappa_1^d/\kappa_2^a$, where the indices $a$ and $d$ refer to the ancilla and data cat qubits of the repetition code, respectively, since $\kappa_1^d$ sets the typical decoherence time of the quantum information to be protected and $1/\kappa_2^a$ sets the typical error syndrome extraction time. By designing an architecture where $\kappa_1^d$ and $1/\kappa_2^a$ are minimized (even if it means increasing $\kappa_1^a$ and $1/\kappa_2^d$), it is, for example, possible to maintain a logical error rate $\epsilon_L = 10^{-7}$ with a repetition code of distance $d=11$ and $\kappa_1/\kappa_2 = 6.3\times10^{-4}$ for an asymmetry $\kappa^a_1/\kappa^d_1 = \kappa^a_2/\kappa^d_2 =20$, instead of $\kappa_1/\kappa_2 = 10^{-4}$ for a symmetric architecture (see Figure 10 of~\cite{LeRegent2023highperformance}).

A second optimization is the use of squeezed cat states $ |\mathcal{C}^\pm_{r, \alpha}\rangle \dot{=}\ S(r)|\mathcal{C}^\pm_{\alpha}\rangle $, where $ S(r) = \exp\left[\frac{r}{2}(a^2-a^{\dagger 2})\right] $ is the squeezing operator. It has been recently theoretically demonstrated~\cite{Schlegel2022,Hillmann2023,Xu2023squeezed} that the scaling of the exponential suppression of bit-flips increases with the squeezing parameter, $ O(e^{-\gamma(r)\bar{n}}) $, where $\gamma(0)\approx 2$ for cat states and $ \gamma(r) > 2 $ for squeezed cat states ($r>0$). Consequently, the targeted bit-flip error probability can be achieved at smaller photon numbers, which allows for an identical physical phase-flip error probability for a higher value of $\kappa_1/\kappa_2$.

A third optimization is the use of an additional Hamiltonian confinement~\cite{Gautier2022}. In this paper, a purely dissipative architecture where the cat states are stabilized by two-photon dissipation at a rate $\kappa_2$ modeled by the Lindbladian operator $D[a^2 - \alpha^2]$ is considered. It has been recently shown how the fidelity and speed of gates on cat qubits can be improved by confining the cats with degenerate Hamiltonians whose ground state corresponds to the code space of the cats~\cite{Xu2021,Putterman2022,Gautier2022,Ruiz2023}. The main difficulty with this approach is to preserve the exponential suppression of bit-flips in the presence of these confinement Hamiltonians~\cite{frattini2022squeezed}.

A potential final optimization  comes from the use of autonomous feedback to correct first-order phase-flip errors induced during gates, which can drastically reduce the gate errors for dissipative cat qubits~\cite{Gautier2023}. Furthermore, with squeezed cat states, this autonomous feedback can also be used to correct errors due to photon loss using the same process.

The study of the precise gains provided by each of these optimizations, and how they can be combined, is out of the scope of this paper.

\subsection{\label{subsec:nothres}Practical fault-tolerance}

As the physical phase-flip error probability of a cat qubit increases linearly with the average photon number $\bar{n}$, the concatenation of cat qubits into phase-flip codes yields an architecture that does not have a threshold: for a fixed value of the error parameter $\kappa_1/\kappa_2$, there are an optimal code distance $d^*$ and an average photon number $\bar{n}^*$ that minimize the logical error rate given by \eqref{eq:logical_error_repcat}. 
Furthermore, some of the circuits used to implement logical gates, while also fault-tolerant, do not have a threshold either~\cite{Guillaud2019,Chamberland2022,Gouzien2023}.
However, for the typical values considered, this minimum is extremely low. For $\kappa_1/\kappa_2 = 10^{-4}$, evaluating \eqref{eq:logical_error_repcat} for $d=81$ and $\bar{n} = 38$ photons gives a logical error rate $\epsilon_L = 10^{-31}$. Thus, it is clear that the absence of a threshold will not be a limiting factor for the logical error rate. Rather, the crucial hypothesis of the architecture is the experimental validity of the exponential suppression of bit-flips. Ultimately, what will set a lower bound on the logical error probability attainable is the typical rate of physical error mechanisms that do not satisfy the error correction assumptions, just like for architectures with a theoretical threshold.

\section{\label{sec:SAT}Efficient code optimisation}

\subsection{\label{subsec:SAT_distance}Efficient computation of the code distance}

Computing the exact distance of a classical code is known to be an NP-hard problem~\cite{vardy1997}. In this work, we relied on two approaches: the first one simply consists in enumerating all of the $2^k-1$ nonzero codewords and computing the minimum Hamming weight. This method becomes very costly for $k \geq 34$, even if cellular automaton codes have the nice advantage that the systematic construction of the codewords increases the computation speed. The second method is to use a SAT solver.

SAT solvers have been optimized to solve the boolean satisfiability problem, \textit{i.e.}\ finding possible values for boolean variables to satisfy a set of constraints. We define these variables and constraints with the generator matrix $G$ of the code, which gives a basis of $k$ codewords, where any codeword $c$ can be written as $ c = \sum^k_{i=1} b_i G_i$. $G_i$ is the codeword corresponding the i-th row of the matrix $G$ and $b_i$ are the boolean variables of our problem.

Given that the goal is to find the minimal Hamming weight of a codeword, we impose the following constraints for a given tested distance $d'$: at least one $b_i$ must be non-zero and the $\sum^k_{i=1} b_i G_i$ must be at most $d'$. If the SAT solver finds a combination of $b_i$ which is a solution, then the distance of the code $d \le d'$, otherwise $d > d'$. The minimal value $d'$ for which the problem is satisfied is the code distance $d$, found using a dichotomic search over $d'$.

Using this method and the z3 SAT solver~\cite{z3}, we were able to compute the exact distance of some codes up to $k=62$. This is possible because the SAT solver does not go through the entire code space but explores the space of possibilities given the constraints in a more subtle way. We numerically observed that the SAT solver is not always faster than the brute force method, however we did not find any simple criteria to predict which method is most efficient for a given code, as SAT solvers are complex algorithms. 

\subsection{\label{subsec:SAT_shape}Code optimisation}

In this section, we detail the procedure to find the optimal stabilizer shapes of cellular automaton codes. The search space consists only of codes with ``pointed'' stabilizer shapes so that any codeword is uniquely determined by its values at the bottom rows of the lattice.

Varying the stabilizer shapes of the code will not change the number of encoded logical qubits for a given width and height but can increase the distance of the code. This can be understood from Figure~\ref{fig:automate} (c-d), showing a codeword of minimal Hamming weight, which consists of a single $\ket{-}$ state in the bottom rows. The Hamming weight, \textit{i.e.}\ the number of $\ket{-}$ states, in this codeword will never exceed the number of qubits within the triangle area. But by changing the stabilizer shapes (\textit{i.e.}\ the rules of the cellular automaton), we may increase the number of $\ket{-}$ states.

We further restrict the search space to a unique stabilizer shape in each row of the lattice. This guarantees that the search remains tractable, while we empirically found that, for smaller instances, removing this constraint does not bring a significant advantage. Brute-forcing the problem becomes quickly intractable, so we used a SAT solver for the largest instances~\cite{z3}. 

We first fix the width and height of the lattice. The boolean variables correspond to the allowed stabilizer shapes (with several boolean variables per shape) and $2^k$ constraints impose that all codewords have a Hamming weight larger than a tested distance $d'$. If the solver is able to find stabilizer shapes which verify all the constraints, it corresponds to a code with distance $d'\ge d$. By progressively increasing $d'$, we are able to find the maximum distance attainable for a cellular automaton code of a given height.

The constraints (\textit{i.e.}\ the Hamming weight of the codewords) need to be expressed with the boolean variables. Thanks to the structure of cellular automaton codes, this task just consists in applying the cellular automaton rules from a given bit-string in the bottom rows conditionally on the boolean variables. This method gives relatively simple expressions compared to a general code where the parity-check matrix would have to be put in normal form. 

\section{\label{sec:bp+osd}BP+OSD}

In this section, we briefly present the BP+OSD decoder and the parameters we chose. This decoder was first introduced in~\cite{panteleev2021} and a more efficient variant of the ordered statistics decoding (OSD) was proposed in~\cite{roffe2020}. The first step of the decoding process is belief propagation where an error probability is assigned to every qubit knowing the syndrome. This marginal probability is calculated iteratively until a fixed maximum number of steps. If the proposed correction is in agreement with the given syndrome, the decoding process stops there. Otherwise, the OSD routine is called. The qubits for which the value is the most reliable at the end of BP are fixed, and a greedy search is performed on the other qubits for the codeword with the smallest Hamming weight and which is in agreement with the syndrome. The OSD order designates the number of qubits involved in this greedy search and the number of combinations tested increases exponentially with the order.

\begin{figure}[t!]
    \centering
    \includegraphics[width=0.8\columnwidth]{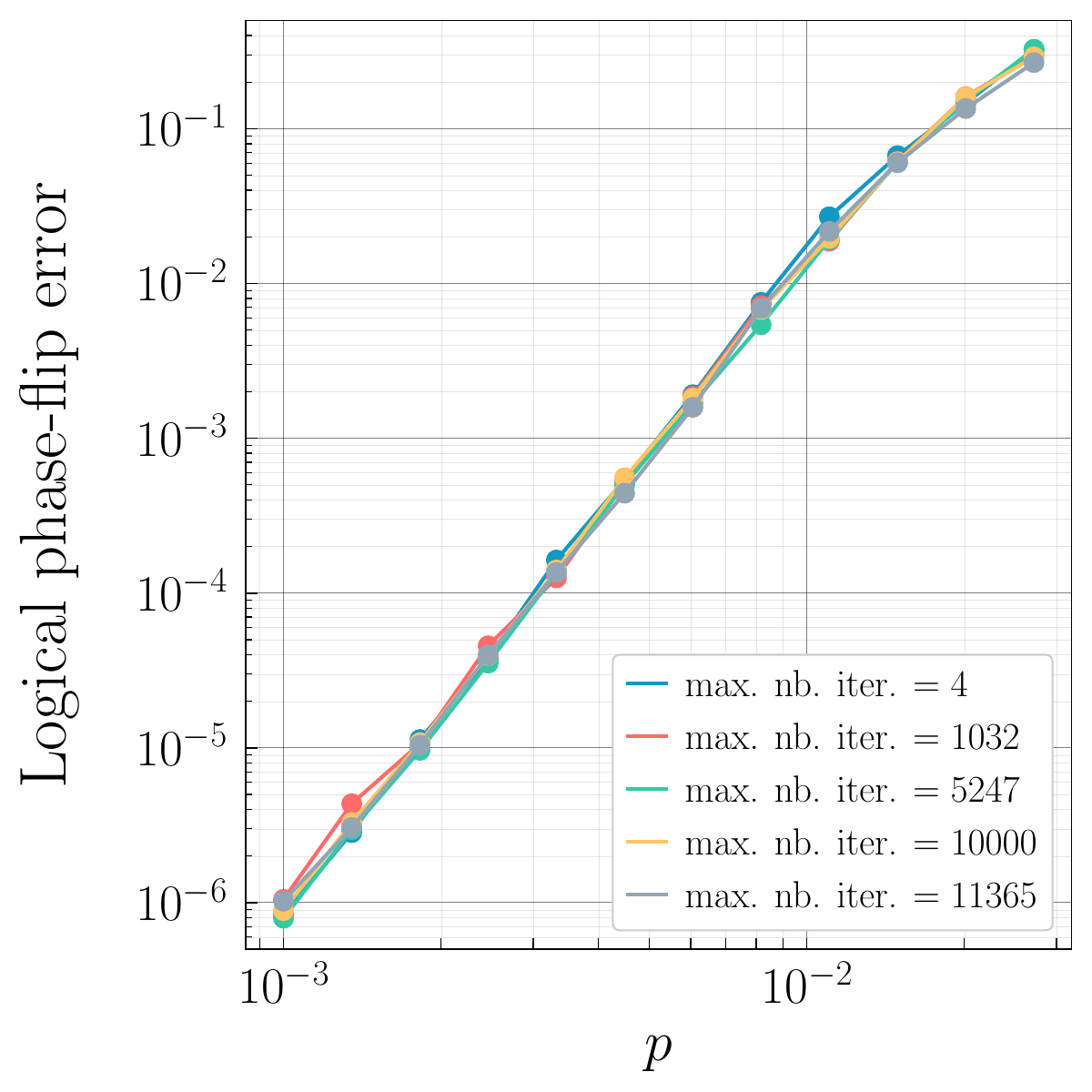}
    \vspace{-0.5cm}
    \caption{\label{fig:bp+osd}
    Logical phase-flip error rate for the cellular automaton code with the stabilizer shape presented in Figure~\ref{fig:automate} (a). The lattice size is set as $H = 3$ and $L=10$ which corresponds to a code of distance 7. The circuit-level noise model is considered and $p$ designates the physical error rate on every gate and idle location in the circuit. The maximum number of iterations of the belief propagation varies showing that this parameter does not play a significant role in the decoding process. The min-sum method is used and the scaling factor is set to 0.625.  
    }
\end{figure}

We chose the ``min-sum'' variant of belief propagation and we have observed no significant difference between the various methods (``product-sum'',``min-sum'', ``product-sum log'' and ``min-sum log'') in the case of the cellular automaton code decoding process. Additionally, the decoding is slightly better when the scaling factor is chosen at 0.625 as proposed in~\cite{panteleev2021} when the ``min-sum'' method is used. Finally, the number of iterations does not appear to play a significant role either (even when set to 4, which corresponds to the cycle length in the Tanner graph of the code) as shown in Figure~\ref{fig:bp+osd}. The numerical results reported in our figures are obtained with a maximum number of iterations of 10,000.

%-------------------------------%
%-------------------------------%

Regarding the OSD routine, increasing the OSD order increases the performance of the decoder but at an exponential cost in time. We confirm, as that was stated in~\cite{roffe2020}, that the ``combination sweep'' method surpasses the ``exhaustive method'' for an equivalent number of configurations tested. We set the OSD order to 60, which gives a number of configurations to search through of 1770. The question of real-time decoding and a possibly faster or more efficient decoder for these very structured codes remains an open question.

\section{\label{sec:state_injection}State injection in the memory layer}

%-------------------------------%
\begin{figure*}[t!]
    \centering
    \includegraphics[width=\textwidth]{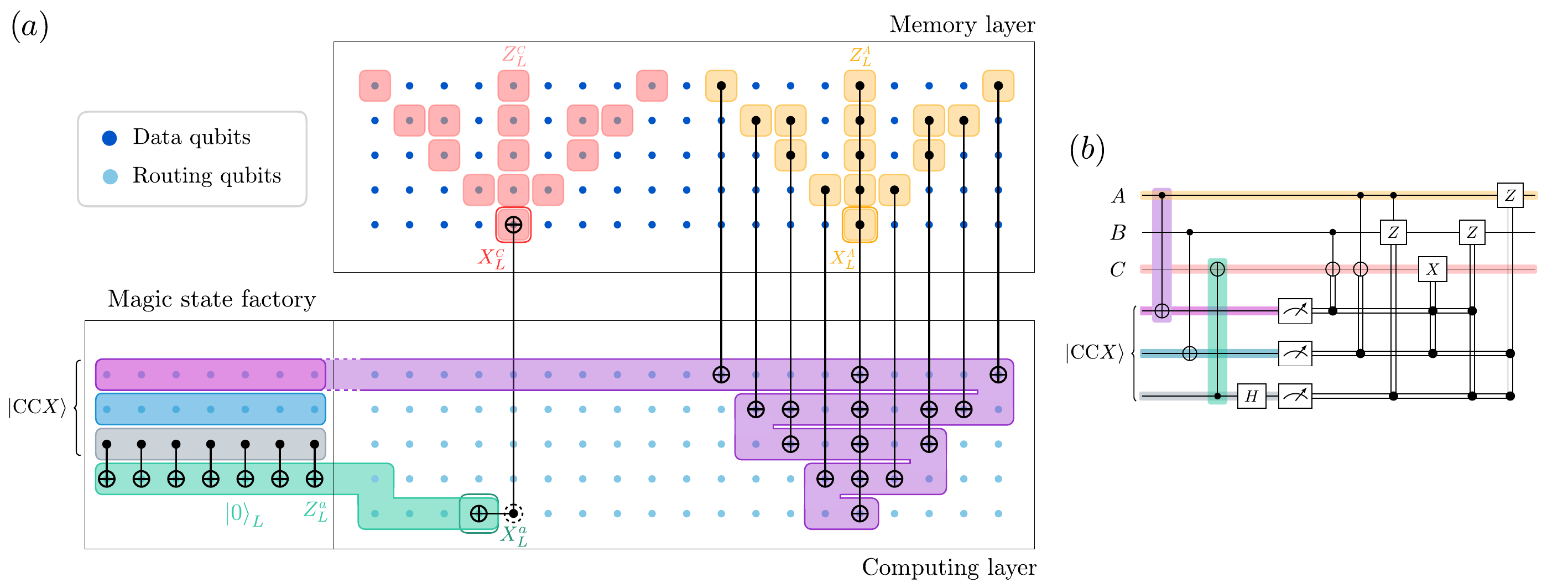}
    \vspace{-0.5cm}
    \caption{\label{fig:Toffoli}
    Injection of the $\ket{\text{CC}X}_L$ magic state. The state is prepared in a dedicated magic state factory, and injected using the logical circuit of Figure (b). As depicted in Figure (a), the implementation requires routing qubits encoded in the repetition code. This routing circuitry can be seen as a second layer on top of the LDPC block, where CNOT are allowed between corresponding qubits in the LDPC and repetition code block. If the magic state is the target (purple), the first step involves ``extending'' or ``moving'' the magic state on the qubits of the computing layer connected to the qubits of the control logical qubit A encoded in the memory layer (yellow). Then, transversal CNOT gates are applied with the logical qubit A as control.
    If the magic state is the control (grey), lattice surgery is used. First, a logical $\ket{0}_L$ is prepared (green). A logical CNOT gate is performed between the control logical magic state and the logical $\ket{0}_L$. This is followed by an $M_{XX}$ measurement between the logical $\ket{0}_L$ and the target logical qubit C (red). This process includes preparing a routing ancilla (dashed circle) in the $\ket{+}$ state, performing two CNOTs with the logical $\ket{0}_L$ and the logical $X_L^C$ of the logical qubit C, and finally measuring the ancilla. The result is the $M_{XX}$ measurement and the procedure is repeated $d$ times to ensure fault-tolerance. Finally, the repetition code logical  ancilla is measured in the $Z$ basis, and Pauli corrections are applied if necessary.
    }
\end{figure*}
%-------------------------------%

\begin{comment}
\subsection{\label{subsec:non_cell_code}Definition of logical Pauli operators for non-cellular automaton codes}

\textit{Pauli operators.}

\textit{Destructive measurement of $X_L$.} In the absence of errors, the value of each logical $X_L$ operator can be obtained by taking the product of the measurement outcomes corresponding to the qubits in their support. Note that, in many cases, including for instance all cellular automaton codes, a logical $X_L$ operator of weight one can be found, which stems from the fact that the phase-flip codes we consider have no bit-flip error correcting capabilities. In a realistic setting with errors, the fault-tolerance of the logical $X_L$ measurement is ensured by taking a majority vote over the values of at least $d$ representatives for each logical $X_L$ operators (equivalent to one another by multiplication by a subset of the stabilizers) with disjoint supports, which values are obtained by multiplication of the relevant $X$ measurement outcomes.
\end{comment}

%\subsection{\label{subsec:magic_state}Fault-tolerant preparation of magic states in repetition codes}

%\subsection{\label{subsec:injection}State injection in the memory layer}

In this section, we detail how magic states, encoded in repetition codes and prepared in dedicated magic state factories, can be injected in the memory layer. We focus on the operations needed in the case of the $\ket{\text{CC}X}_L$ injection, but the schemes can directly be adapted for other magic states. The implementation is shown in Figure~\ref{fig:Toffoli} (a) using the logical injection circuit depicted in Figure~\ref{fig:Toffoli} (b).

The injection of two of the three states of the 3-qubit Toffoli magic state are shown to cover both the case of the magic state being the control and the target of the CNOT with logical qubits in the memory layer. If the magic state is the target, the first step involves ``extending'' or ``moving'' the magic state on the qubits of the computing layer connected to the qubits of the logical qubit encoded in the memory layer. This can be done with well-known stabilizer manipulation techniques~\cite{Fowler2012,Chamberland2022}. Then, transversal CNOT gates are applied with the logical qubit encoded in the memory layer as control. If the magic state is the control, we use lattice surgery as presented in the circuit of Figure~\ref{fig:cnot} (b) with the control being the magic state, a logical repetition code as an ancilla prepared in the state $\ket{0}_L$, and the target being the logical qubit in the memory layer. A transversal logical CNOT gate is performed between the magic state and the ancilla repetition code, followed by a QND logical $X^a_L X^C_L$ measurement and a $Z^a_L$ measurement. Pauli corrections conditioned on the logical measurement outcomes complete the gate teleportation.

\section{\label{sec:RSA_resource_estimation}Factoring 2048-bit RSA integers}

In order to estimate the total number of cat qubits required to factorize a 2048-bit RSA integers, we rely on the implementation of Shor’s algorithm described in~\cite{SangouardPRL2021Factoring2048bit} and adapted to the cat qubit architectures in~\cite{Gouzien2023}, as the set of logical gates is identical.
In this previous work, logical gates are not parallelized, in order to minimize as much as possible the space overhead (number of logical qubits) at the cost of increasing the time overhead (the runtime of the algorithm).
We follow the same implementation, such that the fact that some logical CNOT gates might not be executed in parallel on our codes is not an issue here.

In order to provide a fair comparison of the reduction in overhead, we use the exact same hardware assumptions and error models as in~\cite{Gouzien2023}. We therefore assume a ratio $\kappa_1/\kappa_2 = 10^{-5}$, and an average photon number $\bar n = 21$, in order to reach the extremely low logical error probability $\epsilon_L \approx 10^{-17}$ per logical qubit and per code cycle required to run Shor's algorithm.

Replacing the distance $d=15$ repetition codes by the dense $[24885, 6214, 22]$ code of the $[165 + 8\ell, 34 + 2\ell, 22]$ family in the memory block, the total number of cat qubits in the architecture decreases from $\num{349133}$ cat qubits~\cite{Gouzien2023} to $\num{95941}$ cat qubits (code available here~\cite{code}).
The computation time increases from 4 to 7 days, as a consequence of the weight-4 stabilizers (instead of weight-2) that increases the error correction cycle time from $T_{\rm cycle} = \SI{500}{\nano\second}$ to $T_{\rm cycle} = \SI{900}{\nano\second}$.

\bibliographystyle{unsrtnat}
\bibliography{biblio}% Produces the bibliography via BibTeX.

\end{document}